\documentclass[twocolumn]{aastex61}
\pdfoutput=1 
\usepackage{amsmath,amstext}
\usepackage[T1]{fontenc}
\usepackage{apjfonts} 
\usepackage[figure,figure*]{hypcap}
\usepackage{longtable}
\usepackage{comment}

\newcommand{\HII}{H\,\textsc{ii}}
\newcommand{\HI}{H\,\textsc{i}}
\newcommand{\HeI}{He\,\textsc{i}}
\newcommand{\HeII}{He\,\textsc{ii}}
\newcommand{\NII}{N\,\textsc{ii}}
\newcommand{\OII}{O\,\textsc{ii}}
\newcommand{\OIII}{O\,\textsc{iii}}
\newcommand{\SII}{S\,\textsc{ii}}
\newcommand{\SIII}{S\,\textsc{iii}}

\newcommand{\Hbeta}{H$\beta$}
\newcommand{\threeHe}{$^{3}$He}
\newcommand{\fourHe}{$^{4}$He}
\newcommand{\sevenLi}{$^{7}$Li}

\shorttitle{The PHLEK Survey}
\shortauthors{Hsyu et al.}

\begin{document}

\title{The PHLEK Survey: A New Determination of the Primordial Helium Abundance}

\author{Tiffany Hsyu}
\affiliation{Department of Astronomy \& Astrophysics, University of California Santa Cruz, 1156 High Street, Santa Cruz, CA 95060}
\author{Ryan J. Cooke}
\affiliation{Centre for Extragalactic Astronomy, Department of Physics, Durham University, South Road, Durham DH1 3LE, UK}
\author{J. Xavier Prochaska}
\affiliation{Department of Astronomy \& Astrophysics, University of California Santa Cruz, 1156 High Street, Santa Cruz, CA 95060}
\affiliation{
Kavli Institute for the Physics and Mathematics of the Universe (Kavli IPMU),
5-1-5 Kashiwanoha, Kashiwa, 277-8583, Japan}
\author{Michael Bolte}
\affiliation{Department of Astronomy \& Astrophysics, University of California Santa Cruz, 1156 High Street, Santa Cruz, CA 95060}

\begin{abstract}
We present Keck NIRSPEC and Keck NIRES spectroscopy of sixteen metal-poor galaxies that have pre-existing optical observations. The near-infrared (NIR) spectroscopy specifically targets the \HeI~$\lambda$10830\AA\, emission line, due to its sensitivity to the physical conditions of the gas in \HII\ regions. We use these NIR observations, combined with optical spectroscopy, to determine the helium abundance of sixteen galaxies across a metallicity range $12+\textnormal{log}_{10}(\rm O/\rm H)\,=\,7.13-8.00$.  This data set is combined with two other samples where metallicity and helium abundance measurements can be secured: star-forming galaxies selected from the Sloan Digital Sky Survey spectroscopic database and existing low-metallicity systems in the literature. We calculate a linear fit to these measurements, accounting for intrinsic scatter, and report a new determination of the primordial helium number abundance, $y_{\rm P}\,=\,0.0805^{+0.0017}_{-0.0017}$, which corresponds to a primordial helium mass fraction $Y_{\rm P}\,=\,0.2436^{+0.0039}_{-0.0040}$. Using our determination of the primordial helium abundance in combination with the latest primordial deuterium measurement, $\rm (D/H)_{\rm P}\times10^{5}\,=\,2.527\pm0.030$, we place a bound on the baryon density $\Omega_{\rm b} h^{2}\,=\,0.0215^{+0.0005}_{-0.0005}$ and the effective number of neutrino species $N_{\rm eff}\,=\,2.85^{+0.28}_{-0.25}$. These values are in 1.3$\sigma$ agreement with those deduced from the \textit{Planck} satellite observations of the temperature fluctuations imprinted on the Cosmic Microwave Background.
\end{abstract}


\section{Introduction}
\label{intro}
The abundances of the light elements that were produced during Big Bang Nucleosynthesis (BBN) chiefly depend on: (1) the ratio of the baryon density to photon density, $\eta_{10}\,\equiv\,10^{10}(n_{\rm B}/n_{\gamma})$ and (2) the expansion rate of the Universe \citep{1964Natur.203.1108H, 1966ApJ...146..542P}. Baryonic matter in the Universe just prior to the onset of BBN mostly consisted of free neutrons and protons, which rapidly fused to form deuterium, and subsequently, other light elements. The freeze-out abundances of deuterium and the isotopes of helium and lithium, depend on a competition between the expansion rate of the Universe and the nuclear and weak interaction rates that govern the synthesis of the light elements 
(see the recent BBN reviews by \citealt{2007ARNPS..57..463S, 2012arXiv1208.0032S, 2016RvMP...88a5004C, 2018PhR...754....1P}).


The universal baryon density, $\Omega_{\rm b}h^{2} \simeq \eta_{10}/273.9$ \citep{2006JCAP...10..016S} is determined to $\sim1$ per cent precision via the temperature fluctuations of the Cosmic Microwave Background (CMB). The most recent determination of the baryon density inferred from the CMB is $\Omega_{\rm b}h^{2}\,=\,0.02236\pm0.00016$ (68 per cent confidence limits (CL) of the TT+TE,EE+lowE parameter estimation; see Table 2, Column 4 of \citealt{2018arXiv180706209P}). The expansion rate of the Universe is determined by the total energy density of the Universe. At the time of BBN, the total energy density was dominated by massless and relativistic particles, including photons, electrons, and the three Standard Model neutrinos \citep{2012arXiv1208.0032S, 2017IJMPE..2641001M}. 
The total radiation energy density is parameterized by the effective number of neutrino species, $N_{\rm eff}\,=\,3.046\,+\,\Delta N_{\nu}$ (equivalent to $N_{\nu}\,=\,3\,+\,\Delta N_{\nu}$). For the Standard Model of particle physics and cosmology, $\Delta N_{\nu}\,=\,0$. In the framework of the Standard Model in combination with the \textit{Planck} measurement of $\Omega_{\rm b}h^{2}$, a mean neutron lifetime $\tau_{\rm n}$
, and cross-sections for the relevant reaction rates, the primordial element yields can be predicted to a precision of less than two per cent \citep{2018PhR...754....1P}. 

Similarly, observational measurements of the light element abundances in near-pristine environments provide an opportunity to infer the constituents of the early Universe. These observational measures of the primordial abundances offer an important test of standard Big Bang nucleosynthesis (SBBN); deviations from the SBBN light element abundances would indicate new physics in the early Universe. For example, if $\Delta N_{\nu}\,\neq\,0$, there may be a previously unrecognized particle that changes the total energy density of the Universe and thus the expansion rate of the early Universe (e.g., \citealt{2013JCAP...11..018D}). To assess this possibility, reliable and precise observational measurements of the light element abundances must be made in order to firmly conclude the existence of physics beyond the Standard Model.


The light element nuclides deuterium $\rm D/\rm H$, helium-3 (\threeHe), helium-4 (\fourHe), and lithium-7 (\sevenLi) are made in astrophysically measurable quantities, and have therefore been the targets of historic and current primordial abundance measurements. While all the primordial abundances depend on both the baryon density and the expansion rate of the Universe at the time of BBN, (D/H) and \sevenLi\ are most sensitive to the baryon abundance whereas \fourHe\ is primarily sensitive to the expansion rate of the Universe (see Figure 7 of \citealt{2016RvMP...88a5004C}). \threeHe\ is less sensitive to both the baryon density and the expansion rate than its peer primordial elements but provides orthogonal contours to (D/H) in the $\Delta N_{\nu} - \Omega_{\rm b}h^{2}$ plane \citep{2015ApJ...812L..12C}. A \threeHe\ abundance has been observed and measured in \HII\ regions and planetary nebulae in the Milky Way, but these measures likely do not reflect the primordial \threeHe\ composition, due to contamination by the complicated post-BBN production of \threeHe\ \citep{1995ApJ...444..680O, 2003ApJ...585..611V}.
The primordial abundance of \sevenLi\ can be inferred from the atmospheres of the most metal-poor dwarf stars in our Galaxy. The latest determinations \citep{2009ApJ...698.1803A, 2010A&A...515L...3M, 2010A&A...522A..26S, 2015A&A...582A..74S} are, however, in significant ($\sim6\sigma$) disagreement with the SBBN value \citep{2008JCAP...11..012C, 2011ARNPS..61...47F}, and has been famously dubbed the ``lithium problem''.

The primordial $\rm D/\rm H$ ratio, $(\rm D/\rm H)_{\rm P}$, offers a sensitive probe of the baryon density and has a simple post-BBN chemical evolution. There are no pathways that net produce deuterium, so its abundance should decrease monotonically with increasing metallicity. Currently, the best environments to measure the primordial $\rm D/\rm H$ ratio are high-redshift, near-pristine quasar absorption systems, where the current determination is at the 1 per cent level, $(\rm D/\rm H)_{\rm P}\,=\,(2.527\,\pm\,0.030)\times10^{-5}$, in agreement with SBBN \citep{2018ApJ...855..102C}. 

The mass fraction of \fourHe\ offers a sensitive test of physics beyond the Standard Model \citep{1979ApJ...227..697Y, 1984ApJ...281..493Y, 1981ApJ...246..557O} due to its strong dependence on the effective number of neutrino species. Attempts to measure the primordial \fourHe\ abundance, commonly denoted in the literature by the helium mass fraction, $Y_{\rm P}$, have
most commonly utilized emission line observations of \HII\ regions in low-metallicity dwarf galaxies, defined to have gas phase oxygen abundances less than a tenth solar metallicity, $12+\textnormal{log}_{10}(\rm O/\rm H)\,\leq\,7.69$. 
This method has shown the most promise to reach a $\sim1$ per cent inference on the helium abundance.


\cite{1972ApJ...173...25S} presented an abundance analysis of the extragalactic \HII\ regions I Zwicky 18 (I\,Zw18) and II Zwicky 40 and first suggested that metal-poor systems such as these would be crucial to pin down the primordial helium abundance. Finding new, metal-poor \HII\ regions has historically been difficult, however. While all-sky surveys such as the Sloan Digital Sky Survey (SDSS) have provided a means to identify new, low-metallicity systems \citep{2007ApJ...662...15I, 2007ApJ...665.1115I, 2013A&A...558A..57I, 2017A&A...599A..65G}, the number of metal-poor systems expected from the luminosity function greatly outnumbers the number of known metal-poor systems \citep{2017ApJ...835..159S}. It has been suggested that the most metal-poor systems tend to elude spectroscopic surveys, possibly due to their intrinsically low surface brightnesses as predicted by the luminosity-metallicity relation \citep{2015MNRAS.448.2687J}. Consistent with this line of reasoning, with the exception of the extremely metal-poor but more luminous systems such as I Zw 18 \citep{1972ApJ...173...25S, 1993ApJ...411..655S} and SBS\,0335-052 \citep{1990Natur.343..238I}, 
discoveries of new systems that push on the lowest-metallicity regime have been rare. Yet systems similar to these, i.e., at the hundredth solar metallicity level, are necessary for a precise extrapolation to the primordial helium value. Some recent exceptions include Leo P \citep{2013AJ....146...15G, 2013AJ....146....3S}, AGC\,198691 \citep{2016ApJ...822..108H}, which were both initially found as \HI\ gas rich regions in the Arecibo Legacy Fast ALFA Survey \citep{2005AJ....130.2598G}, the Little Cub \citep{2017ApJ...845L..22H}, J0811$+$4730 \citep{2018MNRAS.473.1956I}, and HSC J1631$+$4426 \citep{2019arXiv191008559K}. Many of the latest efforts to significantly boost the number of low-metallicity \HII\ regions have focused on using photometry to identify candidate systems, followed by spectroscopic confirmation combined with a direct measurement of the metallicity of the system. This method has yielded successful results, with 20~--~60\% of observed systems in these dedicated searches falling in the low-metallicity regime \citep{2015MNRAS.448.2687J, 2017MNRAS.465.3977J, 2017ApJ...847...38Y, 2018ApJ...863..134H, 2019MNRAS.484.1270S}. 


Extracting a measure of the helium abundance of these near-pristine galaxies has its challenges. \HII\ region modelling is believed to suffer from systematic uncertainties (for an incomplete list, see \citealt{2007ApJ...662...15I}) and degeneracies among the model parameters, particularly between the electron density and temperature. This can lead to biases in the determination of the helium abundance (see Figure 3 of \citealt{2015JCAP...07..011A}). To help alleviate these biases, \cite{2014MNRAS.445..778I} included the near-infrared (NIR) \HeI~$\lambda$10830\AA\, line in their helium abundance analysis. The \HeI~$\lambda$10830 line is very sensitive to the electron density, and helps to break the temperature-density degeneracy. \cite{2015JCAP...07..011A} confirmed the importance of \HeI~$\lambda$10830 as an excellent density diagnostic -- the addition of the \HeI~$\lambda$10830 line to their analysis of 11 systems reduced the 1$\sigma$ confidence interval on the electron density by 60 per cent. This reduction of the error on the electron density led to a reduction of the error on the helium abundance of each \HII\ region ranging from 10--80 per cent. 

However, these two works, which have systems in common in their analyses, report primordial helium abundances in mutual disagreement with one another. \cite{2014MNRAS.445..778I} reports $Y_{\rm P}\,=\,0.2551\,\pm\,0.0022$, which is higher than the SBBN predicted value, while \cite{2015JCAP...07..011A} finds $Y_{\rm P}\,=\,0.2449\,\pm\,0.0040$, consistent with the SBBN value of $Y_{\rm P}\,=\,0.24709\,\pm\,0.00017$ \citep{2018PhR...754....1P}. Several other groups have recently reported competitive measurements of the primordial helium abundance in good agreement with the \cite{2015JCAP...07..011A} result, using a range of techniques. For example, \cite{2018MNRAS.478.5301F} use sulphur (S) instead of oxygen (O) as a metallicity tracer, and find that the scatter in the $Y_{\rm P}$ vs $\rm S/\rm H$ plane is reduced compared with $Y_{\rm P}$ vs $\rm O/\rm H$. These authors later employ probabilistic programming methods and find good agreement with their previous work \citep{2019MNRAS.487.3221F}. Other groups have instead focused on modeling a small number of well-selected \HII\ regions to infer the primordial value \citep{2016RMxAA..52..419P, 2019ApJ...876...98V}. It is perhaps promising that the different data sets used and the different modelling approaches employed yields mostly consistent results (with the exception of the value reported by \citealt{2014MNRAS.445..778I}). However, it is still necessary to take caution of confirmation bias (see e.g. Figure 8 of \citealt{2012arXiv1208.0032S}), and understand why models are currently unable to simultaneously reproduce all of the observed \HI\ and \HeI\ emission lines of some \HII\ regions. 

Motivated by the dearth of metal-poor systems that push on the lowest-metallicity regime and the need for more high-quality, complementary optical and NIR spectra of external galaxies, we conducted a dedicated survey to identify new, metal-poor systems via SDSS photometry \citep{2018ApJ...863..134H}. Our follow-up spectroscopic survey of 94 objects found almost half of them to be in the low-metallicity regime, and our findings included one of the lowest-metallicity systems currently known, the Little Cub \citep{2017ApJ...845L..22H}. After initial metallicity estimates, we obtained spectroscopy of a subset of the most promising systems, with a focus on obtaining high signal-to-noise ($S/N$) optical and NIR spectra. In this paper, we use this new sample, along with some previous systems in the literature, to report a new determination of the primordial helium abundance.



In Section \ref{obs_and_data}, we describe the details of the full sample of galaxies that we use in this paper. This includes our own sample of new complementary optical and NIR data, for which we also include details of the observations, data reduction, and integrated emission line flux measurements. We supplement our data set with galaxies from the SDSS spectroscopic database and the HeBCD sample from \citet{2004ApJ...602..200I, 2007ApJ...662...15I}.
The components of our model and the subsequent MCMC analysis used to solve for the best fit parameters of our \HII\ regions are described in Section \ref{model}. 
In Section \ref{analysis}, we assess the potential systematics and select the most reliable set of \HII\ regions to use in our determination of the primordial helium abundance. We discuss the implications of our work and consider future improvements to primordial helium work, both in observations of new systems and in model enhancements, in Section \ref{discussion}. Finally, we summarize our main conclusions in Section \ref{conclusion}.

\section{Data Compilation and Preparation} 
\label{obs_and_data}
A well constrained measurement of the primordial helium abundance requires accurate measurements of the oxygen and helium abundance from a sizeable sample of galaxies that span a range of metallicities. In this section, we describe the observations of our galaxy sample, which populates the lowest-metallicity regime. Throughout this paper, we refer to our galaxy sample as the Primordial Helium Legacy Experiment with Keck (PHLEK) sample. We supplement our PHLEK sample with existing spectra from SDSS and the \citet{2004ApJ...602..200I, 2007ApJ...662...15I} HeBCD data set. This combined sample provides a set of measurements that cover a broad range of metallicity. We note that the three data sets that make up our final, full sample of galaxies are thus likely heterogeneous data sets, and the degree of our involvement in processing each sample (e.g., converting the two-dimensional, raw data into integrated emission line fluxes, varies).

\subsection{Keck Observations}
\label{oursample}
The primary goal of our observational program is to increase the sample size of very metal-poor galaxies where reliable oxygen and helium abundances can be determined. To this end, we acquired optical and near-infrared spectra of metal-poor \HII\ regions in nearby dwarf galaxies using Keck Observatory, requiring that the spectra have confident detections of:
\begin{itemize}
    \item the temperature sensitive [\OIII]~$\lambda$4363\AA\, line for a direct measurement of the oxygen abundance
    \item at least five optical \HeI\ emission lines to reliably determine the physical state of the \HII\ regions, including: \HeI~$\lambda$3889\AA\,, $\lambda$4026\AA\,, $\lambda$4471\AA\,, $\lambda$5015\AA\,, $\lambda$5876\AA\,, $\lambda$6678\AA\,, and $\lambda$7065\AA
    \item the NIR \HeI~$\lambda$10830\AA\, line, whose emissivity is the most sensitive \HeI\ emission line to the density of the gas, relative to P$\gamma~\lambda$10940\AA
\end{itemize}
In addition to these emission lines, we also detect in our spectra the [\OII] doublet at $\lambda\lambda$3727,\,3729\AA $ $, the [\OIII] doublet at $\lambda\lambda$4959,\,5007\AA $ $, the [\NII] doublet at $\lambda\lambda$6548,\,6584\AA $ $, the [\SII] doublet at $\lambda\lambda$6717,\,6731\AA $ $, and the Balmer series from H$\alpha$ to at least H8.

To ensure that we observe the same region of each system either on multiple nights or on different instruments, we acquire each target by first centering on a bright nearby star, then applying an offset to the target based on SDSS astrometry. Additionally, we requested that our optical and near-infrared nights be allocated within a week of one another such that our complementary observations for a given target be at similar airmass and parallactic angle. For the observations, we matched the slit widths of different instruments as best as possible. Spectroscopic observations of our metal-poor galaxy sample took place during semesters 2015B, 2016A, and 2018A (program IDs:  U052LA/U052NI, U091LA/U091NS, U172).

\subsubsection{Optical Spectroscopy}
\label{optical_observations}
Optical spectroscopic observations of 32 metal-poor systems were made using the Low Resolution Imaging Spectrometer (LRIS) with the atmospheric dispersion corrector (ADC) at the W.M. Keck Observatory. LRIS has separate blue and red channels. On the blue side, our setup utilized the 600/4000 grism, which has an unbinned dispersion of 0.63~\AA\,pix$^{-1}$. On the red side, we used the 600/7500 grating, which has an unbinned dispersion of 0.8~\AA\,pix$^{-1}$. Using this instrument setup, the D560 dichroic, and a long slit, the full wavelength coverage achieved is $\sim$3200--8600~\AA, with the separate blue and red channels covering $\sim$3200--5600~\AA\, and $\sim$5400--8600~\AA, respectively. We use 2$\times$2 binning during readout. The blue and red channels have nominal FWHM resolutions of 2.6~\AA\, and 3.1~\AA\, for our adopted $0.70^{\prime\prime}$ slit.
While the separate blue and red arms overlap in wavelength coverage, we take caution about the accuracy of the measurements here, as data near the region of overlap is compromised by the dichroic.

Our spectra were taken with a $175\times\,0.70^{\prime\prime}$ slit, oriented at the parallatic angle.  Our total exposure times range from 3\,$\times$\,1200~s to 3\,$\times$\,1800~s. We obtained bias frames, arc frames, and dome flats at the beginning of the night. For wavelength calibration on the blue side, we observed Hg, Cd, and Zn arc lamps; on the red side, we observed Ne, Ar, and Kr arc lamps. Photometric standard stars G191B2B, BD$+$284211, Feige 34, Feige 66, Feige 110, and/or HZ44 were observed at the start and end of each night for flux calibration. Excluding five previously unreported systems which are presented here, our observed and derived physical properties of the galaxies based on Keck+LRIS spectra are reported in \cite{2018ApJ...863..134H}.

\subsubsection{Near-Infrared Spectroscopy}
\label{NIR_observations}
We acquired complementary NIR observations for 16 of our 32 galaxies with optical spectroscopy. NIR observations were made using NIRSPEC in semesters 2015B and 2016A and the Near-Infrared Echellette Spectrometer (NIRES) in 2018A. Our NIRSPEC observations were done in low resolution mode using the NIRSPEC-1 filter, which offers a wavelength coverage of $\sim$9470--12100\AA. NIRES covers wavelengths $\sim$9400--24500\AA\, across five orders, with a gap between 18500--18800\AA, though this wavelength gap does not affect our observation goals.


Our NIRSPEC observations were made using the 42$\times\,0.72^{\prime\prime}$ slit to best match the slit width of our LRIS observations. The NIRES slit is fixed at 18$\times\,0.55^{\prime\prime}$. We observed all targets with the slit oriented at the parallactic angle. All NIR observations were made using an ABBA nod pattern for exposure times of 8\,$\times$\,250~s to 8\,$\times$\,360~s each. We obtained dome flats at the beginning of each night. An A0V calibration star near each of our science targets was observed following each observation for flux calibration.


\subsection{Data Reduction}
For optical LRIS observations, the two-dimensional raw images were individually bias subtracted, flat-field corrected, cleaned for cosmic rays, sky-subtracted, extracted, wavelength calibrated, and flux calibrated, all using \textsc{PypeIt} (previously \textsc{Pypit}), a Python based spectroscopic data reduction package.\footnote{\textsc{PypeIt} is available from: \url{http://doi.org/10.5281/zenodo.3506873}} We used a boxcar extraction technique to extract a single one-dimensional (1-D) spectrum of each object.\footnote{Optimal extraction methods are unsuitable here due to the extended nature of our systems.} Multiple observations of the same target were coadded by weighting each exposure by the inverse variance at each pixel.

\begin{figure*}
\includegraphics[width=1.0\textwidth]{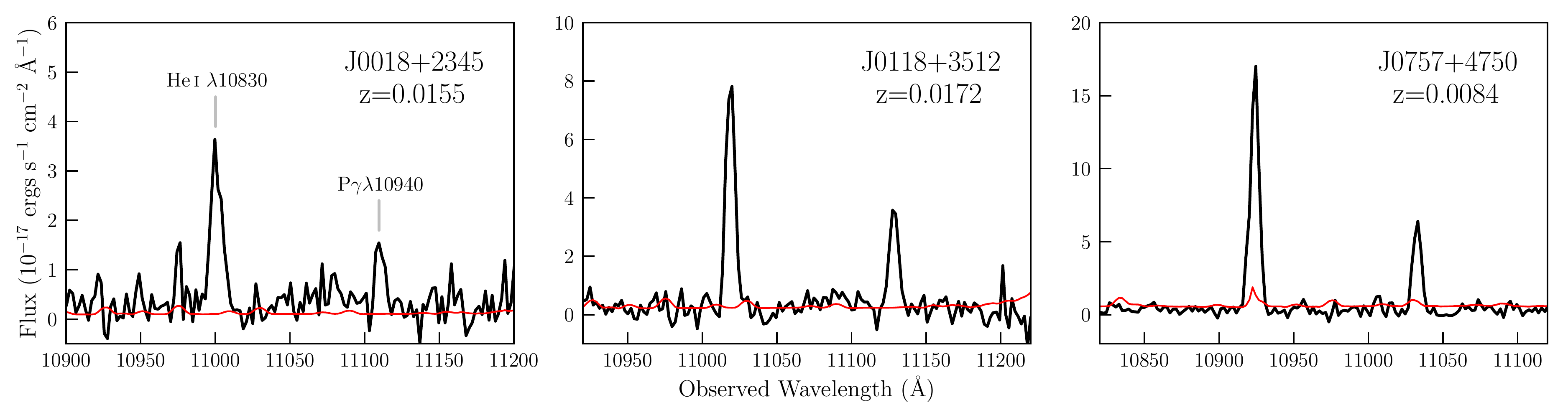}
\caption{The coadded near-infrared spectra (shown in black) of the first three systems listed in Table \ref{table:nir_fluxes}, as collected using NIRSPEC at Keck Observatory. The error spectra are shown in red. Only a small window of NIRSPEC's entire $\sim$9470--12100\AA\, wavelength range is shown in these panels to best highlight the relevant emission lines of interest, \HeI~$\lambda$10830\, and P$\gamma~\lambda$10940, which are marked in the left panel.}
\label{fig:nir_spec}
\end{figure*}

For our NIR data, \textsc{PypeIt} combines a single set of ABBA observations during the reduction as A$+$A\,$-$\,(B$+$B), yielding an extracted 1-D spectrum at nod location A. Similarly, the frames are combined as B$+$B\,$-$\,(A$+$A) for a spectrum at nod location B. \textsc{PypeIt} first flat-fields the individual frames, then combines and subtracts relevant frames, which removes the bias level and performs a first order sky subtraction. \textsc{PypeIt} wavelength calibrates using the OH sky lines. Flux calibrations for NIR observations are performed separately from the automated reduction routine using the \texttt{pypeit\_flux\_spec} script. Our NIR observations of each target were acquired in two sets of ABBA observations, such that the final coadded spectrum consists of four 1-D extracted spectra, comprised of two spectra of A$+$A\,$-$\,(B$+$B) and two spectra of B$+$B\,$-$\,(A$+$A). We show an example of our reduced and coadded NIR spectra in Figure \ref{fig:nir_spec}.

\subsection{SDSS Sample}
\label{sdss}
In addition to our new sample of metal-poor systems observed at Keck, we also use the SDSS spectroscopic database to identify additional emission-line galaxies that can be included in our primordial helium work. The SDSS sample complements our PHLEK sample described in Section \ref{oursample} by providing a sample of higher metallicity galaxies. It also offers the potential to significantly increase the number of systems available for helium abundance analyses.

To take advantage of this database, we queried the SDSS \texttt{specObj} database for systems that are suitable to our analysis. Our query requires that the systems are: (1) classified as starburst galaxies and (2) within a redshift range of 0.02 $ < z < $ 0.15, such that the [\OII] doublet and \HeI~$\lambda$7065 lines, necessary for a metallicity and helium abundance, fall on the detector. Our SQL query can be found in Appendix \ref{CasJobsquery}.

For the resulting galaxies, we calculate the emission line fluxes using the method described in Section \ref{emlinefit} and filtered the systems to keep those with confident detections of: (1) the temperature sensitive [\OIII] $\lambda$4363\AA\, line for a direct metallicity, and (2) multiple \HeI\ lines, to measure the helium abundance. We impose these criteria using the following $S/N$ cuts, where $S/N$ is defined to be the measured $F(\lambda)/\sigma(F(\lambda))$:
\begin{gather*} 
S/N([\textnormal{\OIII}]~\lambda4363) \geq 5 \\
S/N(\textnormal{\HeI~$\lambda$5876}) \geq 20 \\
S/N(\textnormal{\HeI~$\lambda$4471}) \geq 3 \\
S/N(\textnormal{\HeI~$\lambda$6678}) \geq 3 \\
S/N(\textnormal{\HeI~$\lambda$7065}) \geq 3
\end{gather*}
Of these \HeI\ lines, the \HeI~$\lambda$5876 line is typically the most significantly detected. We therefore require the strongest $S/N$ condition on this line to ensure a confident detection of the weaker \HeI\, lines.

These steps filtered the SDSS spectroscopic database down to 1053 candidate systems to be included in our analysis. For reference, the peak of the metallicity distribution of this SDSS sample is $(\rm O/\rm H)\times10^{5}\,=\,13.24$, whereas the peak of the metallicity distribution of our PHLEK galaxies, including the systems presented in \citet{2018ApJ...863..134H} and here, is $(\rm O/\rm H)\times10^{5}\,=\,4.82$. These values correspond to $12+\textnormal{log}_{10}(\rm O/\rm H)$ values of 8.12 and 7.68, respectively.
  
\subsection{Emission Line Flux Measurements}
\label{emlinefit}
For the Keck and SDSS samples, we calculate the integrated emission line fluxes by summing the total flux above the continuum level at each emission line, where the continuum level and its error are modelled using the Absorption LIne Software (\textsc{ALIS}, see \citealt{2014ApJ...781...31C} for a more detailed description of the software).\footnote{\textsc{ALIS} is available at: \url{https://github.com/rcooke-ast/ALIS}} ALIS simultaneously fits the emission line profile using a Gaussian model and the surrounding continuum using a 1- or 2-D Legendre polynomial and determines the best fit parameters of the Gaussian and continuum model using a $\chi^{2}$ minimization approach. Systems with high emission line fluxes, however, are not well-represented by a single Gaussian. We therefore adopt the continuum model and its associated error from the ALIS output, and use this to inform our calculation of the total flux above the continuum level. The width of the emission line included in the integrated flux is set to be $\pm5$ pixels around the closest pixel to the redshifted central wavelength of the emission line. Two exceptions are the [\OII] doublet, which has a width of $\pm7$ pixels to encompass the full width of the blended doublet, and \HeI~$\lambda$5015 where we take only 3 pixels ($\sim$1.9\AA) to the left of the central wavelength to avoid contamination from the [\OIII] $\lambda$5007 line (we still use 5 pixels to the right). We map the pixels to an array of change in wavelength at each pixel, $d\lambda_i$, and determine the integrated flux:
\begin{equation}
\ F(\lambda)\,=\, \sum_{i} \, (F_{i}-h_{i})\,d\lambda_{i}
\label{integrated_flux}
\end{equation}
where $F_{i}$ is the flux and $h_{i}$ is the continuum level.

The integrated flux measurements of our optical Keck spectra are published in \cite{2018ApJ...863..134H}, except for five new systems, which are listed in Table \ref{table:optical_fluxes}. Our Keck NIR observations are described in Table \ref{table:nir_fluxes}. The measured emission line flux ratios of our systems, along with the 1053 systems derived from the SDSS galaxy sample that satisfy our $S/N$ criteria, are also available on GitHub as MCMC input files as part of our primordial helium code,  \textsc{yMCMC}.\footnote{\textsc{yMCMC} is available at: \url{https://github.com/tiffanyhsyu/yMCMC}}

\begin{deluxetable*}{cccccc}
\tablewidth{0.99\textwidth}
\tablecaption{Optical emission line fluxes of \HII\ regions in our Primordial Helium Legacy Experiment with Keck} 
\tablehead{
\colhead{} & \colhead{} & \colhead{} & \colhead{Target Name} & \colhead{} & \colhead{}}
\startdata 
Ion & J0118$+$3512 & J0757$+$4750 & J1204$+$5259 & J1214$+$1245 & J1322$+$5425 \\
\hline
{[O\,\textsc{ii}]\,$\lambda$3727+3729} & 0.8494\,$\pm$\,0.0040 & 0.6092\,$\pm$\,0.0025 & 1.113\,$\pm$\,0.011 & 1.578\,$\pm$\,0.012 & 0.4346\,$\pm$\,0.0028 \\
{H8+He\,\textsc{i}\,$\lambda$3889} & 0.1464\,$\pm$\,0.0022 & 0.1656\,$\pm$\,0.0017 & 0.1495\,$\pm$\,0.0038 & 0.1376\,$\pm$\,0.0070 & 0.1836\,$\pm$\,0.0023 \\
{He\,\textsc{i}\,$\lambda$4026} & 0.0107\,$\pm$\,0.0016 & 0.0154\,$\pm$\,0.0010 & 0.0129\,$\pm$\,0.0055 & \nodata & 0.0163\,$\pm$\,0.0012 \\
{H$\delta$\,$\lambda$4101} & 0.2149\,$\pm$\,0.0023 & 0.2242\,$\pm$\,0.0016 & 0.1901\,$\pm$\,0.0063 & 0.199\,$\pm$\,0.011 & 0.2350\,$\pm$\,0.0023 \\
{H$\gamma$\,$\lambda$4340} & 0.4198\,$\pm$\,0.0026 & 0.4217\,$\pm$\,0.0018 & 0.3703\,$\pm$\,0.0068 & 0.438\,$\pm$\,0.011 & 0.4441\,$\pm$\,0.0026 \\
{[O\,\textsc{iii}]\,$\lambda$4363} & 0.0640\,$\pm$\,0.0016 & 0.0906\,$\pm$\,0.0012 & 0.0686\,$\pm$\,0.0050 & 0.0421\,$\pm$\,0.0094 & 0.0753\,$\pm$\,0.0013 \\
{He\,\textsc{i}\,$\lambda$4472} & 0.0335\,$\pm$\,0.0015 & 0.03741\,$\pm$\,0.00098 & 0.0251\,$\pm$\,0.0048 & 0.0420\,$\pm$\,0.0096 & 0.0325\,$\pm$\,0.0010 \\
{He\,\textsc{ii}\,$\lambda$4686} & 0.0321\,$\pm$\,0.0020 & \nodata & \nodata & \nodata & 0.01052\,$\pm$\,0.00081 \\
{H$\beta$\,$\lambda$4861} & 1.0000\,$\pm$\,0.0035 & 1.0000\,$\pm$\,0.0024 & 1.0000\,$\pm$\,0.0088 & 1.000\,$\pm$\,0.012 & 1.0000\,$\pm$\,0.0035 \\
{[O\,\textsc{iii}]\,$\lambda$4959} & 1.0207\,$\pm$\,0.0036 & 1.3272\,$\pm$\,0.0024 & 1.599\,$\pm$\,0.011 & 0.786\,$\pm$\,0.011 & 0.9812\,$\pm$\,0.0032 \\
{[O\,\textsc{iii}]\,$\lambda$5007} & 3.0626\,$\pm$\,0.0058 & 4.1087\,$\pm$\,0.0040 & 4.679\,$\pm$\,0.021 & 2.112\,$\pm$\,0.015 & 2.9386\,$\pm$\,0.0050 \\
{He\,\textsc{i}\,$\lambda$5015} & 0.0260\,$\pm$\,0.0015 & 0.0139\,$\pm$\,0.0011 & 0.0217\,$\pm$\,0.0042 & 0.0094\,$\pm$\,0.0091 & 0.02444\,$\pm$\,0.00091 \\
{He\,\textsc{i}\,$\lambda$5876} & 0.1149\,$\pm$\,0.0043 & 0.03479\,$\pm$\,0.00041 & 0.1282\,$\pm$\,0.0051 & 0.0649\,$\pm$\,0.0062 & 0.0840\,$\pm$\,0.0014 \\
{H$\alpha$\,$\lambda$6563} & 3.3499\,$\pm$\,0.0057 & 0.9684\,$\pm$\,0.0010 & 3.472\,$\pm$\,0.011 & 2.5969\,$\pm$\,0.0095 & 2.6708\,$\pm$\,0.0080 \\
{[N\,\textsc{ii}]\,$\lambda$6584} & 0.0391\,$\pm$\,0.0018 & 0.01066\,$\pm$\,0.00026 & 0.0423\,$\pm$\,0.0041 & 0.0257\,$\pm$\,0.0056 & 0.01532\,$\pm$\,0.00082 \\
{He\,\textsc{i}\,$\lambda$6678} & 0.0307\,$\pm$\,0.0017 & 0.00983\,$\pm$\,0.00026 & 0.0297\,$\pm$\,0.0040 & 0.0300\,$\pm$\,0.0054 & 0.02376\,$\pm$\,0.00096 \\
{[S\,\textsc{ii}]\,$\lambda$6717} & 0.1101\,$\pm$\,0.0029 & 0.02792\,$\pm$\,0.00031 & 0.1472\,$\pm$\,0.0042 & 0.1305\,$\pm$\,0.0055 & 0.04456\,$\pm$\,0.00092 \\
{[S\,\textsc{ii}]\,$\lambda$6731} & 0.0852\,$\pm$\,0.0018 & 0.02070\,$\pm$\,0.00029 & 0.1125\,$\pm$\,0.0041 & 0.1114\,$\pm$\,0.0064 & 0.03147\,$\pm$\,0.00100 \\
{He\,\textsc{i}\,$\lambda$7065} & 0.0321\,$\pm$\,0.0020 & 0.01005\,$\pm$\,0.00025 & 0.0353\,$\pm$\,0.0037 & 0.0269\,$\pm$\,0.0076 & 0.01970\,$\pm$\,0.00086 \\
\hline
$F$(H$\beta$) ($\times$10$^{-17}$\,erg\,s$^{-1}$\,cm$^{-2}$) & 332.9\,$\pm$\,1.2 & 780.3\,$\pm$\,1.8 & 81.51\,$\pm$\,0.71 & 75.17\,$\pm$\,0.91 & 542.2\,$\pm$\,1.9 \\
\hline \hline
\enddata 
\tablecomments{Optical emission line fluxes of systems observed using LRIS and previously unreported in \cite{2018ApJ...863..134H}. The reported values are integrated flux measurements given relative to the H$\beta$ flux, which is also quoted for reference. These fluxes are uncorrected for reddening, since reddening is a parameter we later solve for in our MCMC analysis.} 
\label{table:optical_fluxes}
\end{deluxetable*}

\begin{deluxetable*}{cccc}[ht]
\tablewidth{12pt}
\tablecaption{Near-infrared emission line fluxes of \HII\ regions in our Primordial Helium Legacy Experiment with Keck} 
\tablehead{ 
\colhead{Galaxy} & \colhead{$F$(He\,\textsc{i}\,$\lambda$10830)} & \colhead{$F$(P$\gamma$)} & \colhead{$F$(He\,\textsc{i}\,$\lambda$10830)/$F$(P$\gamma$)} 
}
\startdata 
J0018$+$2345 & 29.59\,$\pm$\,0.90 & 10.96\,$\pm$\,0.73 & 2.699\,$\pm$\,0.082 \\
J0118$+$3512 & 62.0\,$\pm$\,1.7 & 26.9\,$\pm$\,1.4 & 2.301\,$\pm$\,0.061 \\
J0757$+$4750 & 120.1\,$\pm$\,6.4 & 46.4\,$\pm$\,4.6 & 2.59\,$\pm$\,0.14 \\
KJ5 & 45.4\,$\pm$\,2.9 & 11.2\,$\pm$\,2.7 & 4.06\,$\pm$\,0.26 \\
KJ5B & 31.3\,$\pm$\,1.4 & 14.6\,$\pm$\,1.3 & 2.142\,$\pm$\,0.093 \\
J0943$+$3326 & 10.76\,$\pm$\,0.99 & 2.33\,$\pm$\,0.73 & 4.61\,$\pm$\,0.42 \\
Little Cub & 8.3\,$\pm$\,2.1 & 4.6\,$\pm$\,3.1 & 1.81\,$\pm$\,0.46 \\
J1204$+$5259 & 52.0\,$\pm$\,2.4 & 23.7\,$\pm$\,2.4 & 2.20\,$\pm$\,0.10 \\
KJ97 & 28.3\,$\pm$\,2.5 & 6.7\,$\pm$\,1.7 & 4.26\,$\pm$\,0.38 \\
KJ29 & 42.2\,$\pm$\,3.0 & 17.2\,$\pm$\,3.8 & 2.45\,$\pm$\,0.18 \\
J1322$+$5425 & 91.4\,$\pm$\,5.3 & 38.7\,$\pm$\,4.1 & 2.36\,$\pm$\,0.14 \\
KJ2 & 68.9\,$\pm$\,2.1 & 14.7\,$\pm$\,1.5 & 4.68\,$\pm$\,0.14 \\
J1655$+$6337 & 116.6\,$\pm$\,9.9 & 31.8\,$\pm$\,9.3 & 3.66\,$\pm$\,0.31 \\
J1705$+$3527 & 25.06\,$\pm$\,0.38 & 8.32\,$\pm$\,0.67 & 3.011\,$\pm$\,0.046 \\
J1757$+$6454 & 52.7\,$\pm$\,1.3 & 20.1\,$\pm$\,1.1 & 2.623\,$\pm$\,0.063 \\
J2213$+$1722 & 380\,$\pm$\,14 & 190\,$\pm$\,10 & 2.003\,$\pm$\,0.072 \\
\enddata 
\tablecomments{Observed near-infrared emission line flux and emission line flux ratios of 16 galaxies observed using NIRSPEC or NIRES at Keck Observatory. The fluxes are integrated flux measurements in units of 10$^{-17}$\,erg\,s$^{-1}$\,cm$^{-2}$ and not corrected for reddening, which is a parameter we solve for in the MCMC.} 
\label{table:nir_fluxes}
\end{deluxetable*}

The total reported error of the emission line fluxes comprises of two terms added in quadrature: the measured error of the integrated emission line flux and an assumed 2 percent relative flux uncertainty to account for the error of the flux calibration. The latter follows a common procedure in primordial helium work \citep{1994ApJ...431..172S, 2007ApJ...662...15I} and is taken from \cite{1990AJ.....99.1621O}, which quantified the absolute flux uncertainties on a set of 25 standard stars now recognized as the Hubble Space Telescope spectrophotometric standards. \cite{1990AJ.....99.1621O} found these standard stars to be reliable to about 1--2\% across the optical wavelength regime (see Table VI of \citealt{1990AJ.....99.1621O}).

\subsection{HeBCD Sample}
\label{hebcd}
To the PHLEK and SDSS samples, we also add the HeBCD sample of galaxies from \cite{2004ApJ...602..200I, 2007ApJ...662...15I}, a fraction of which have follow-up NIR observations reported in \cite{2014MNRAS.445..778I}. Their sample consists of 93 total systems, 21 of which have unique optical plus NIR spectroscopy, i.e., we do not consider systems with optical spectra reported for multiple regions but one NIR spectra. This is to ensure that the optical and NIR emission line fluxes originate from observations of the same part of a singular \HII\ region. The HeBCD data set have metallicities that overlap with both our PHLEK sample and the SDSS sample, with a median metallicity of $(\rm O/\rm H)\times10^{5}\,=\,9.40$ or $12+\textnormal{log}_{10}(\rm O/\rm H)\,=\,$7.97. For these systems, we take the reported emission line flux ratios and equivalent widths but re-determine their best-fit parameters, including the helium abundance, using our model, as described below in Section \ref{model}. Updated optical data of the HeBCD sample were obtained from E. Aver (2018; private communication) and include the \HeI~$\lambda$4026 flux and corrections to the original values found in \cite{2007ApJ...662...15I}.


\section{Model Overview}
\label{model}
Most of the hydrogen and helium in an \HII\ region is in an ionized state. Thus, the number abundance ratio of helium to hydrogen, $y$, of an \HII\ region is given by the sum of the abundance ratios of singly and doubly ionized helium:
\begin{equation}
\ y\,=\,\frac{\textnormal{He}^{+}}{\textnormal{H}^{+}}~+~\frac{\textnormal{He}^{++}}{\textnormal{H}^{+}}\,=\,y^{+}~+~y^{++}
\label{equation:helium_abundance}
\end{equation}
The $y^{+}$ and $y^{++}$ abundances depend on the intrinsic helium to hydrogen ratio of the \HII\ region, along with the detailed physical state of the ionized gas and the surrounding stellar population. Since the observed \HeI\ and \HI\ relative line ratios depend on these physical parameters, we can measure the \HeI\ and \HI\ line ratios to pin down the physical conditions of the ionized gas. Our analysis follows a similar approach to that described first by \cite{2011JCAP...03..043A} and subsequently by \cite{2012JCAP...04..004A, 2013JCAP...11..017A, 2015JCAP...07..011A}.

Our code \textsc{yMCMC} solves for the best fit parameters that reproduce the measured emission line ratios of our sample of galaxies described in Section \ref{obs_and_data}. \textsc{yMCMC} closely follows the model and methods mentioned in the above works using a Python implementation of a Markov Chain Monte Carlo (MCMC) sampler, \textsc{emcee} \citep{2013PASP..125..306F}, to survey an 8-dimensional parameter space:
\begin{itemize}
    \item the ionized helium abundance, $y^{+}$
    \item the electron temperature, $T_{\rm e}$ [K]
    \item the electron density, $n_{\rm e}$ [cm$^{-3}$]
    \item the reddening parameter, $c(\rm H\beta)$
    \item the underlying hydrogen stellar absorption, $a_{\rm H}$ [\AA], \\ normalized to the amount of absorption at H$\beta$
    \item the underlying helium stellar absorption, $a_{\rm He}$ [\AA], \\ normalized to the amount of absorption at \HeI~$\lambda$4026
    \item the helium optical depth parameter, $\tau_{\rm He}$, \\ normalized to the value at \HeI~$\lambda$3889
    \item the ratio of neutral to singly ionized hydrogen density, $\xi\,\equiv\,n(\textrm{H\,\textsc{i}})/n(\textrm{H\,\textsc{ii}})$
\end{itemize}
At each step of the MCMC chain, our model predicts the \HeI\ and \HI\ emission line fluxes as a ratio relative to \Hbeta\, and calculates the log-likelihood function of the model:

\begin{equation}
\textnormal{log}(\mathcal{L})\,=\,\sum_{\lambda}\frac{\big(\frac{F(\lambda)}{F(\rm H\beta)}_{\rm p}\,-\,\frac{F(\lambda)}{F(\rm H\beta)}_{\rm m}\big)^{2}}{\sigma(\lambda)^{2}} \\
\label{equation:chi_squared}
\end{equation}
where $\sigma(\lambda)$ is the uncertainty of the flux ratio of each emission line. The subscripts $p$ and $m$ represent the predicted and measured flux ratios, respectively. The predicted flux ratio of the hydrogen emission lines is given by:
\begin{multline}
\label{equation:fluxH}
\frac{F(\lambda)}{F(\rm H\beta)}_{\rm p}\,=\,\frac{E(\lambda)}{E(\rm H\beta)}\frac{\frac{EW({\rm H}\beta)~+~a_{\rm H}(\rm H\beta)}{EW(\rm H\beta)}}{\frac{EW(\lambda)~+~a_{\rm H}(\lambda)}{EW(\lambda)}} \\
\times\,\frac{1+\frac{C}{R}(\lambda)}{1+\frac{C}{R}(\rm H\beta)}10^{-f(\lambda)~c(\rm H\beta)}
\end{multline}
Here, $E(\lambda)$ is the emissivity of an emission line at wavelength $\lambda$, $EW(\lambda)$ is the measured equivalent width ($EW$) of the emission line, $\frac{C}{R}(\lambda)$ is the collisional to recombination correction factor, and $f(\lambda)$ is the reddening law. These individual components are discussed in further detail below. For helium emission lines, the predicted flux ratio is similarly given by:

\begin{multline}
\label{equation:fluxHe}
\frac{F(\lambda)}{F(\rm H\beta)}_{\rm p}\,=\,y^{+}\frac{E(\lambda)}{E(\rm H\beta)}\frac{\frac{EW({\rm H}\beta)~+~a_{\rm H}(\rm H\beta)}{EW(\rm H\beta)}}{\frac{EW(\lambda)~+~a_{\rm He}(\lambda)}{EW(\lambda)}}f_{\tau}(\lambda) \\
\times\,\frac{1+\frac{C}{R}(\lambda)}{1+\frac{C}{R}(\rm H\beta)}10^{-f(\lambda)~c(\rm H\beta)}
\end{multline}
where $f_{\tau}(\lambda)$ is the optical depth function. 

As shown in Equations \ref{equation:fluxH} and \ref{equation:fluxHe}, our model for predicting flux ratios depends on the measured quantity $EW(\rm H\beta$), which has a corresponding uncertainty. To account for this uncertainty, at each step of the MCMC chain, we draw a new value for $EW(\rm H\beta$) from a Gaussian distribution with a width equal to the measured uncertainty. This is the same approach adopted by \cite{2011JCAP...03..043A}. Additionally, we perturb $EW(\rm H\alpha)$ and $EW(\rm P\gamma)$ for our PHLEK sample and for systems with NIR data, respectively. In these two cases, we require $EW(\rm H\alpha)$ and $EW(\rm P\gamma)$ to predict the theoretical $F(\rm P\gamma$)~/~$F(\rm H\beta$) and $F(\rm H\alpha$)~/~$F(\rm H\beta$) ratios, which we use to match our predicted model fluxes to the format of our measured input fluxes (see Section \ref{MCMC} for details).

We further note that the equivalent width and the measured flux are not independent of one another. However, a conserved quantity is the height of the continuum around each emission line, $h(\lambda$). To ensure that the equivalent widths used in Equations \ref{equation:fluxH} and \ref{equation:fluxHe} scale appropriately with the predicted fluxes, we introduce the following relation:

\begin{equation}
\label{equation:flux_EW_cont}
\frac{F(\lambda)}{F(\rm H\beta)}\,=\,\frac{EW(\lambda)}{EW(\rm H\beta)}\,\frac{h(\lambda)}{h(\rm H\beta)}
\end{equation}
which allows us to rewrite Equations \ref{equation:fluxH} and \ref{equation:fluxHe}, removing $EW(\lambda$) entirely, as follows:
\begin{multline}
\label{equation:fluxH_noEW}
\frac{F(\lambda)}{F(\rm H\beta)}_{\rm p}\,=\,\frac{E(\lambda)}{E(\rm H\beta)}\frac{EW({\rm H}\beta)~+~a_{\rm H}(\rm H\beta)}{EW(\rm H\beta)} \\ \times\,\frac{1+\frac{C}{R}(\lambda)}{1+\frac{C}{R}(\rm H\beta)}10^{-f(\lambda)~c(\rm H\beta)} - \frac{a_{\rm H}(\lambda)}{EW(\rm H\beta)}\frac{h(\lambda)}{h(\rm H\beta)}
\end{multline}
\begin{multline}
\label{equation:fluxHe_noEW}
\frac{F(\lambda)}{F(\rm H\beta)}_{\rm p}\,=\,y^{+}\frac{E(\lambda)}{E(\rm H\beta)}\frac{EW({\rm H}\beta)~+~a_{\textnormal{H}}(\rm H\beta)}{EW(\rm H\beta)}f_{\tau}(\lambda) \\
\times\,\frac{1+\frac{C}{R}(\lambda)}{1+\frac{C}{R}(\rm H\beta)}10^{-f(\lambda)~c(\rm H\beta)} - \frac{a_{\rm He}(\lambda)}{EW(\rm H\beta)}\frac{h(\lambda)}{h(\rm H\beta)}
\end{multline}

With Equations \ref{equation:fluxH_noEW} and \ref{equation:fluxHe_noEW}, \textsc{yMCMC} generates the model flux ratios given a set of parameters drawn from the MCMC. Motivated by physically meaningful limits, we impose the following uniform priors on the following parameters:
\begin{gather*} 
0.06 \leq y^{+} \leq 0.10 \\
0 \leq \textnormal{log}_{10}(n_{\rm e}/\rm cm^{-3}) \leq 3 \\
0 \leq c(\rm H\beta) \leq 0.5 \\
0 \leq a_{\rm H} \leq 10 \\
0 \leq a_{\rm He} \leq 4 \\
0 \leq \tau_{\rm He} \leq 5 \\
-6 \leq \textnormal{log}_{10}(\xi) \leq -0.0969
\end{gather*}
The upper limit placed on $\textnormal{log}_{10}(\xi)$ here is unrealistic for an \HII\ region, as this upper bound would imply that only 55 per cent of the gas in the \HII\ region is ionized. We allow our MCMC to explore this regime, but disqualify systems that have best recovered solutions that are unreasonable for \HII\ regions (see Section \ref{qualification}).

To ensure that the electron temperature parameter explored by our MCMC stays within reasonable limits for the system, we include the following weak prior on $T_{\rm e}$:
\begin{equation}
\textnormal{log}(p)\,=\,-\frac{\chi^{2}}{2}-\frac{(T_{\rm e}-T_{\rm m})^{2}}{2\sigma^{2}}
\end{equation}
where we take $\sigma$ to be 0.2$T_{\rm m}$ and $T_{\rm m}$ is the direct measurement of the electron temperature based on the [\OIII]~$\lambda$4363~/ ~($\lambda$4959\,+\,$\lambda$5007) ratio. This weak prior was also implemented by \cite{2011JCAP...03..043A}, who demonstrated with synthetic data that the above prior improves the recovery of the input model parameters and removes local minima near the edges of the likelihood distributions. We also require that the electron temperature is within the range $10,000~\rm K\,\leq$ $T_{\rm e}\,\leq\,22,000~\rm K$. In the following sections, we describe in detail the implementation of each term in Equations \ref{equation:fluxH_noEW} and \ref{equation:fluxHe_noEW}.

\subsection{Emissivity}
\label{emissivity}
The \HI\ and \HeI\ emissivities, denoted by $E(\lambda$), provide a measure of the energy released per unit volume and time. $E(\lambda)$ is expressed in units of erg\,s$^{-1}$\,cm$^{-3}$ throughout.

\subsubsection{Hydrogen Emissivity}
\label{hydrogen_emissivity}
Our model determines the emissivity of an \HI\, line at a given temperature and density following the hydrogen emissivity calculations made by P. Storey (2018; private communication) assuming Case B recombination. The Storey 2018 emissivities extend the \cite{2015MNRAS.446.1864S} hydrogen emissitivies down to the lowest density regime explored by our model, log$_{10}(n_{\rm e}/\rm cm^{-3}$)\,=\,0, and are available up to log$_{10}(n_{\rm e}/\rm cm^{-3}$)\,=\,5 at log$_{10}(n_{\rm e}/\rm cm^{-3}$)\,=\,1 intervals. The emissivities are calculated over the temperature range $T_{\rm e}$\,=\,5,000~--~25,000~K, at 1,000~K intervals. In our model, we interpolate linearly within this temperature and density grid using \textsc{SciPy}'s \texttt{RectBivariateSpline()}.

The implementation of \HI\ emissivities in our model assumes no error in the emissivity value. As an estimate of the uncertainty on these emissivities, we compare $E(\rm H\beta)$ from Storey 2018 with the parameterization of $E(\rm H\beta)$ by R. L. Porter (given in Eq. 3.1 of \citealt{2010JCAP...05..003A}; we note that this parameterization is independent of the electron density). Within the temperature and density ranges of interest, the emissivities differ by 0.10--0.55 per cent. At a fixed temperature, the difference in emissivities increases with increasing electron density. The ratio of the H$\alpha$, H$\gamma$, and H$\delta$ to H$\beta$ emissivities from Storey 2018 differ by 0.10--0.20 percent compared to the parameterizations in \citet{2010JCAP...05..003A}. That is, the extended Storey 2018 \HI\ emissivities are not expected to significantly change our model and therefore the resulting best fit MCMC parameters. Rather, they make the \HI\ emissivity grid more self-consistent, as it no longer relies on extrapolations to the lowest density regime.

\subsubsection{Helium Emissivity}
\label{helium_emissivity}
The \HeI\, line emissivity at a given temperature and density is determined in a similar manner to the \HI\, emissivities. Our model adopts the \HeI\, emissivities introduced in \cite{2013JCAP...11..017A}, which project the \cite{2012MNRAS.425L..28P, 2013MNRAS.433L..89P} \HeI\, emissivities onto a finer grid. The \cite{2012MNRAS.425L..28P, 2013MNRAS.433L..89P} emissivities assume Case B recombination and are calculated for a grid of temperatures ranging from $T_{\rm e}$\,=\,10,000~K~--~25,000~K and densities from log$_{10}(n_{\rm e}/\rm cm^{-3}$)\,=\,1~--~5. We linearly interpolate the \HeI\, emissivities within this temperature and density grid using \textsc{SciPy}'s \texttt{RectBivariateSpline()} interpolator.


\subsection{Collisional to Recombination Ratio, $\frac{C}{R}(\lambda)$}
\label{CR}
The collisional to recombination ratio, $\frac{C}{R}(\lambda)$, corrects for the amount of neutral hydrogen and helium atoms excited to higher energy states due to collisions with electrons, and the emission detected as a result of the electrons subsequently cascading down to lower energy levels.

\subsubsection{Hydrogen}
\label{hydrogen_CR}
Following the method of calculating the collisional to recombination correction factor in \cite{2010JCAP...05..003A}, the $\frac{C}{R}(\lambda)$ ratio of an \HI\, line is given by:

\begin{equation}
\begin{split}
\ \frac{C}{R}(\lambda) &\,=\,\frac{n(\textnormal{H\,\textsc{i}})~(\sum\limits_{i}q_{1\rightarrow i}BR_{i\rightarrow j})~BR_{j\rightarrow n}n_{\rm e}}
{n(\textnormal{H\,\textsc{ii}})~\alpha_{+\rightarrow j}BR_{j\rightarrow n}n_{\rm e}} \\
& =\xi\,\times\,\frac{\sum\limits_{i}q_{1\rightarrow i}BR_{i\rightarrow j}}
{\alpha_{+\rightarrow j}}
\end{split}
\label{equation:CRcorrection}
\end{equation}

Here, $n(\textnormal{H\,\textsc{i}})$ and  $n(\textnormal{H\,\textsc{ii}})$ are, respectively, the neutral and ionized hydrogen densities in units of cm$^{-3}$. The ratio of these densities is defined as $\xi$ and solved for as one of the free parameters in the MCMC.

The subscripts used in the numerator of Equation \ref{equation:CRcorrection} represent energy level transitions to: the energy level $i$, which is above or equal to the transition level of interest, $j$ (i.e., $i\geq j$). In the denominator of Equation \ref{equation:CRcorrection}, the effective recombination rate from an ionized energy level above energy level $j$ is given by $\alpha_{+\rightarrow j}$. The subsequent downward transition from $j\rightarrow n=2$ then gives rise to the Balmer wavelength of interest. For the Paschen series, the transition of interest becomes $j\rightarrow n=3$.

The numerator of Equation \ref{equation:CRcorrection} expresses the contribution of emission stemming from collisional excitations. $q_{1\rightarrow i}$ represents the rate coefficient of collisional excitation from the ground state $n=1$ to a higher energy level $i$, in cm$^{3}~$s$^{-1}$. The value of $q_{1\rightarrow i}$ depends on the effective collision strength of the transition, $\Upsilon_{1i}$, reported in \cite{653091858, 653091857} such that:
\begin{equation}
\ q_{1\rightarrow i}\,=\,4.004\,\times\,10^{-8}\,\sqrt{\frac{1}{k_\textnormal{B}T}}\,\textnormal{exp}\big(\frac{-13.6 \textnormal{eV}(1-\frac{1}{i^{2}})}{k_\textnormal{B}T}\big)\Upsilon_{1i} \;\; .
\label{equation:q1i}
\end{equation}
Once collisionally excited to the higher energy level $i$, the electron can cascade downward via various paths leading to energy level $j$. The probability of the different transition paths,
$n'l'\rightarrow nl$, are expressed as branching ratios, $BR_{i\rightarrow j}$, and reported in \cite{1983ADNDT..28..215O}. The $j\rightarrow 2$ transition of interest then occurs, with various path probabilities captured in the term $BR_{j\rightarrow 2}$. Finally, the collisional contribution to emission depends both on the density of neutral hydrogen atoms and the density of electrons in the gas available for collisions, $n(\textnormal{H\,\textsc{i}})$ and $n_{\rm e}$.

\cite{653091858, 653091857} only report collision strengths up to principle quantum number $n=5$. While the contribution of collisional emission for transitions $n>5$ is expected to be small, we apply scaling factors to quantify the collisional contributions of emission lines emanating from transitions $n>5$, specifically H$\delta$ ($6\rightarrow 2$), H$8$ ($8\rightarrow 2$), and P$\gamma$ ($6\rightarrow 3$):

\begin{equation}
\ \frac{C}{R}(\lambda)\,=\,\frac{C}{R}(\textnormal{H}\gamma|\textnormal{P}\beta)~\textnormal{exp} \big(\frac{-13.6\textnormal{eV}(\frac{1}{5^{2}}-\frac{1}{j^{2}})}{k_\textnormal{B}T} \big)
\label{equation:CRscale}
\end{equation}

For all other transitions $n\leq5$, the $n'l'$ orbitals we include are:

\begin{itemize}
    \item H$\alpha$: $3s,~3p,~3d,~4s,~4p,~4d,~4f$
    \item H$\beta$: $4s,~4p,~4d,~4f,~5s,~5p,~5d,~5f,~5g$
    \item H$\gamma$: $5s,~5p,~5d,~5f,~5g$
\end{itemize}

The denominator of Equation \ref{equation:CRcorrection} represents the amount of emission from the recombination of free electrons with ionized hydrogen atoms and the subsequent cascade down to less excited states. This is expressed by $\alpha_{+\rightarrow j}$, the rate that an ionized hydrogen atom recombines and transitions from higher energy levels, represented by $+$, down to the $j$ energy level, in units of cm$^{3}~$s$^{-1}$. 
However, the value of $\alpha_{+\rightarrow j}$ must be proportional to all the emission that subsequently emanates from transitions \textit{out} of energy level $j$: 

\begin{equation}
\ n_{+}n_{\rm e}\alpha_{+\rightarrow j}h\nu\,=\,\sum\limits_{k}E(\lambda)_{j\rightarrow k}
\label{equation:emis_to_recomb}
\end{equation}
And therefore, we can use emissivities of the subsequent transitions out of an energy level $j$ as a proxy for $\alpha_{+\rightarrow j}$. For example, any recombination that brings an electron to energy level $j=4$ will then transition out of $j=4$ via either the $4\rightarrow 3$ transition or the $4\rightarrow 2$ transition. 
Thus, rather than using values of the recombination rates in the literature, we choose to substitute $\alpha_{+\rightarrow j}$ with our latest hydrogen emissivities following Equation \ref{equation:emis_to_recomb}. This allows us to take advantage of the more refined temperature and density grid for which we have emissivity values.

\begin{figure}[!ht]
\includegraphics[width=0.5\textwidth]{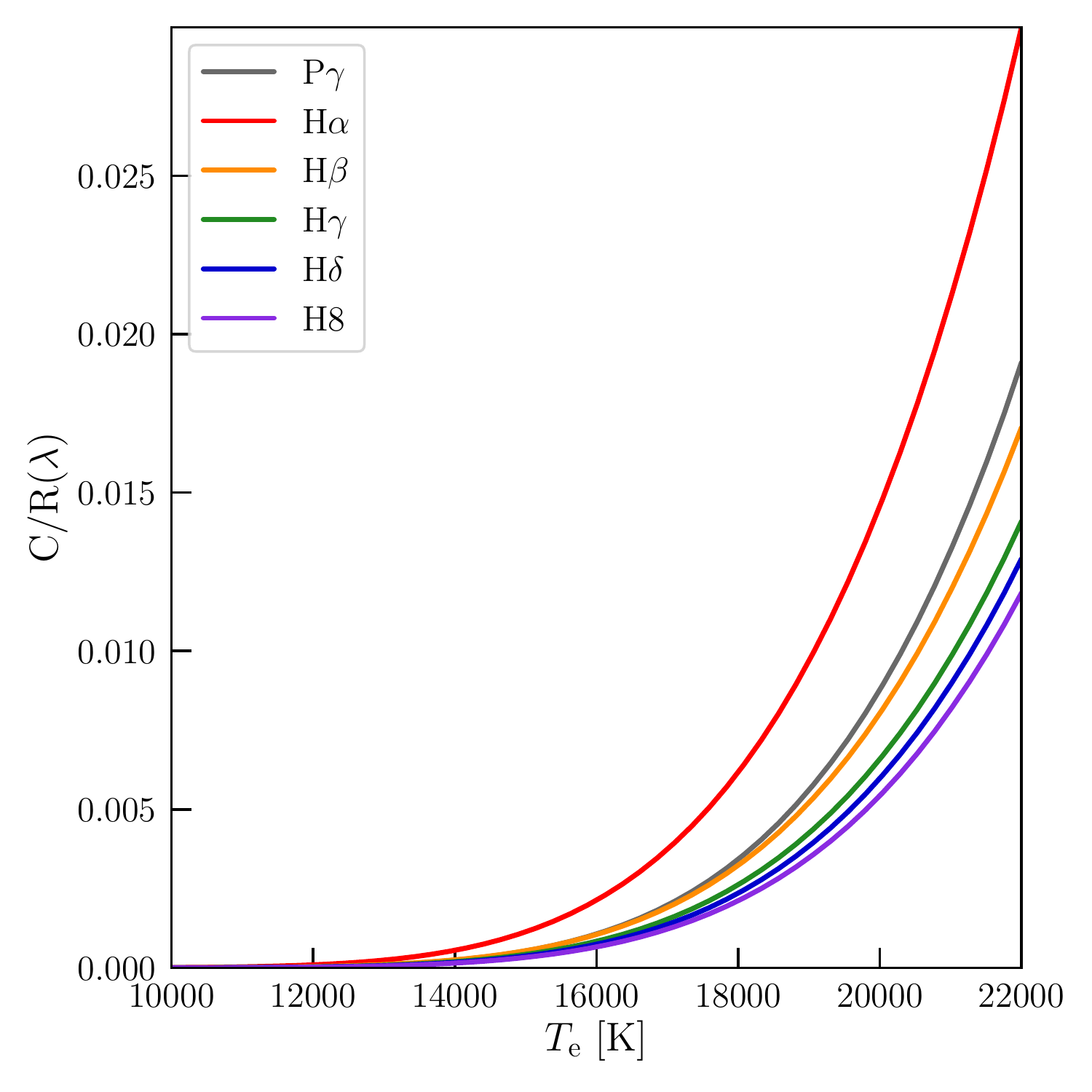}
\caption{The collisional correction of observed \HI\ lines as a function of temperature. The correctional factor calculates the amount of observed emission due to the collisional excitation of neutral hydrogen. In this figure, we use a value of $n$(\HI)~/~$n$(\HII)$\,\equiv\,\xi\,=\,10^{-4}$. The corrections for H$\delta$ and H8 are scaled from the correction for H$\gamma$ following Equation \ref{equation:CRscale}, and similarly, the correction for P$\gamma$ is scaled from P$\beta$.}
\label{fig:hydrogenCR}
\end{figure}

The functional form of the collisional corrections for our \HI\ lines of interest, over a range of temperatures, are shown in Figure \ref{fig:hydrogenCR}. In this figure, we use a neutral to ionized hydrogen density ratio of $\xi\,=\,10^{-4}$ for illustration.

\subsubsection{Helium}
\label{helium_CR}
The $\frac{C}{R}(\lambda)$ correction for \HeI\, is folded in to the \textsc{cloudy} modelling done by \cite{2012MNRAS.425L..28P, 2013MNRAS.433L..89P} in their latest emissivity work. The correctional factors are therefore included in our implementation of the interpolated \HeI\ emissivities described in Section \ref{helium_emissivity}. We refer readers to Section 3 of \cite{2013JCAP...11..017A} for a more detailed description of the collisional contribution included in the \cite{2012MNRAS.425L..28P, 2013MNRAS.433L..89P} emissivities. We note that since the \citeauthor{2013MNRAS.433L..89P} emissivities ($E(\lambda)/E(\rm H\beta)$ in Equation \ref{equation:fluxHe_noEW}) include the collisional correction, we set the value of $\frac{C}{R}(\lambda)\,=\,0$ in the version of Equation \ref{equation:fluxHe_noEW} implemented in our code \textsc{yMCMC}. 


\subsection{Underlying Absorption}
\label{stellar_absorption}
The observed \HI\ and \HeI\ emission line fluxes are compromised by underlying stellar absorption from the atmospheres of the stars in the \HII\ region. Failing to correct for the missing emission can lead to underestimating the total integrated flux of the \HI\ and \HeI\ lines. The amount of underlying absorption depends on the particular stellar population in the galaxy. However, information about the specific stellar population, its age, and its metallicity, along with the possibility of multiple stellar populations, is difficult to extract from long-slit spectroscopy of the \HII\ region. While older works assumed a constant $EW$ of underlying stellar absorption at all \HI\ and \HeI\ lines \citep{b9890a982907415fa4d3eb85829e9388}, it is now recognized that these values are wavelength dependent; assuming a constant amount of underlying absorption across the spectrum biases the derived value of the primordial helium abundance. Various works have estimated the average amount of stellar absorption expected at each \HI\ and \HeI\ line based on synthetic spectra \citep{1999ApJS..125..489G, 2005MNRAS.357..945G}.

Our wavelength dependent underlying absorption corrections are given as coefficients normalized to the amount of absorption present at H$\beta$ for \HI\ lines and \HeI~$\lambda$4471 for \HeI\ lines. Our model incorporates the coefficient values introduced by 
\cite{2010JCAP...05..003A} and repeated in Equations 4.2 and 4.3 of \cite{2015JCAP...07..011A} to include the stellar absorption at the NIR \HeI~$\lambda$10830\, and P$\gamma$\, lines. The \cite{2010JCAP...05..003A} values represent the relative $EW$s of underlying absorption suitable over a range of stellar ages as calculated from a suite of stellar population models. These coefficients are summarized in the following subsections.

\subsubsection{Hydrogen}
\label{hydrogen_absorption}
The coefficients of underlying hydrogen stellar absorption are given below, normalized to the amount of absorption present at H$\beta$, referenced as the variable $a_{\rm H}(\lambda)$, and in units of $EW$ (\AA). The value at $a_{\rm H}(\rm H8)$ is extrapolated from a linear fit to the wavelength and coefficients from H$\beta$ to H$\delta$. We exclude H$\alpha$ from the fit due to the decreasing nature of the underlying absorption at redder wavelengths.

\begin{align}
    a_\textnormal{H}(\rm H\alpha)&=0.942 & a_\textnormal{H}(\rm H\delta)=0.896 \\
    a_\textnormal{H}(\rm H\beta)&=1.000 & a_\textnormal{H}(\rm H8)=0.882 \nonumber \\
    a_\textnormal{H}(\rm H\gamma&)=0.959 & a_\textnormal{H}(\rm P\gamma)=0.400 \nonumber
\end{align}

\subsubsection{Helium}
\label{helium_absorption}
We apply a correction to the optical and NIR \HeI\ emission lines to account for underlying stellar absorption. The following values are given in $EW$ (\AA) and normalized to the amount of underlying helium absorption at \HeI~$\lambda$4471, denoted by the general variable $a_{\rm He}(\lambda)$. That is, the amount of stellar absorption at a given \HeI\ line is the correctional coefficient at its wavelength, multiplied by $a_{\rm He}(\lambda)$. The value of $a_{\rm He}(\textnormal{\HeI}~\lambda5015)$ given is determined using a linear fit to the coefficients of all other listed optical \HeI\ lines.

\begin{align}
    a_\textnormal{He}(\textnormal{\HeI}~\lambda3889)&=1.400 & a_\textnormal{He}(\textnormal{\HeI}~\lambda5876)=0.874 \\
    a_\textnormal{He}(\textnormal{\HeI}~\lambda4026)&=1.347 & a_\textnormal{He}(\textnormal{\HeI}~\lambda6678)=0.525 \nonumber \\
    a_\textnormal{He}(\textnormal{\HeI}~\lambda4471)&=1.000 & a_\textnormal{He}(\textnormal{\HeI}~\lambda7065)=0.400 \nonumber \\
    a_\textnormal{He}(\textnormal{\HeI}~\lambda5015)&=1.016 & a_\textnormal{He}(\textnormal{\HeI}~\lambda10830)=0.800 \nonumber
\end{align}

\subsection{Reddening Correction}
\label{reddening}
Our observed emission line fluxes are expected to suffer from reddening due to dust along the line-of-sight. The theoretical emissivities of the \HI\ recombination lines are well-known and relatively insensitive to the temperature and density of the gas, and are therefore well-suited to the determination of the amount of reddening present in the observed spectrum. To correct for this effect, we include a logarithmic correction factor $c(\rm H\beta)$ in Equations \ref{equation:fluxH_noEW} and \ref{equation:fluxHe_noEW}. When combined with a reddening law, $f(\lambda)$, the amount of extinction as a function of wavelength can be inferred. In our work, we assume the reddening law presented in Equations 2 and 3 of \cite{1989ApJ...345..245C}. Using the formulation given by \cite{1989ApJ...345..245C}, we generate a list of $f(\lambda)$ values for a wavelength grid of 1000 values between 3100~--~13000\AA. We then linearly interpolate this functional form at the observed wavelengths of the \HI\ and \HeI\ emission lines.

The best fit value of $c(\textnormal{H}\beta)$ includes reddening within our own Milky Way and in the observed system. For our sample of galaxies, however, the reddening correction is expected to be small because our candidate systems were selected to be away from the disk of the Milky Way and are expected to be of lower metallicity, where the effects of dust are less important. We note that it is typical to assume no error in the assumed reddening law \citep{2000astro.ph..7081O}.

\subsection{Optical Depth Function}
\label{optical_depth}
The optical depth function is a correction term that accounts for photons that are emitted but subsequently re-absorbed or scattered out of our line of sight. Accordingly, the correction depends on optical depth, the temperature, and the density of the gas. We use a set of optical depth corrections that are suited to the modelling of low metallicity \HII\ regions \citep{2002ApJ...569..288B}. These assume Case B recombination, a spherically symmetric \HII\ region with no systemic expansion or velocity gradients, and are valid for a temperature and density range of $T_{\rm e}$\,=\,12,000~--~20,000~K and $n_{\rm e}$\,=\,1~--~300~cm$^{-3}$. The coefficients of the fits to the optical depth correction are presented in Table 4 of \cite{2002ApJ...569..288B} and can be found listed in Equation A3 in the Appendix of \cite{2004ApJ...617...29O}. The formulation of \HeI~$\lambda$10830 is not included in the original work, but we apply the formula given by Equation 2.2 of \cite{2015JCAP...07..011A}. For completeness, we give the functional form of the fits below in Equation \ref{equation:opticaldepth}, and the coefficients of individual \HeI\, lines are given in Table \ref{table:opticaldepth}.
\begin{equation}
\ f_{\tau}(\lambda)\,=\,1+\frac{\tau}{2}[a~+~(b_0~+~b_1n_{\rm e}~+~b_2n_{\rm e}^{2})~T_{4}]
\label{equation:opticaldepth}
\end{equation}
where $T_{4}\,=\,T_{\rm e}/10,000~$K.

\begin{deluxetable*}{ccccc}[ht]
\tablewidth{12pt}
\tablecaption{Coefficients of the optical depth function} 
\tablehead{ 
\colhead{Wavelength (\AA)} & \colhead{$a$} & \colhead{$b_{0}$} & \colhead{$b_{1}$} & \colhead{$b_{2}$}
}
\startdata 
3889 & $-1.06\times10^{-1}$ & 5.14$\times10^{-5}$ & $-4.20\times10^{-7}$ & 1.97$\times10^{-10}$ \\
4026 & 1.43$\times10^{-3}$ & 4.05$\times10^{-4}$ & 3.63$\times10^{-8}$	& \nodata \\
4471 & 2.74$\times10^{-3}$ & $0.81\times10^{-4}$ & $-1.21\times10^{-6}$ & \nodata \\ 
5015 & 0.0 & 0.0 & 0.0 & 0.0 \\
5876 & 4.70$\times10^{-3}$ & 2.23$\times10^{-3}$ & $-2.51\times10^{-6}$ & \nodata \\
6678 & 0.0 & 0.0 & 0.0 & 0.0 \\
7065 & 3.59$\times10^{-1}$ & $-3.46\times10^{-2}$ & $-1.84\times10^{-4}$ & 3.039$\times10^{-7}$ \\  
10830 & 1.49$\times10^{-2}$ & 4.45$\times10^{-3}$ & $-6.34\times10^{-5}$ & 9.20$\times10^{-8}$ \\
\enddata
\tablecomments{Coefficients of the optical depth correction factor that appear in Equation \ref{equation:opticaldepth}. This functional form has been developed specifically for helium abundance measurements of \HII\ regions and are valid only in the temperature and density range of $T_{\rm e}$\,=\,12,000~--~20,000~K and $n_{\rm e}$\,=\,1~--~300~cm$^{-3}$. There are no optical depth corrections for the singlet lines \HeI~$\lambda$5015 and \HeI~$\lambda$6678, i.e., $f_{\tau}(\lambda)\,=\,1$.} 
\label{table:opticaldepth}
\end{deluxetable*}


\subsection{MCMC Details}
\label{MCMC}
To determine the best fit parameters of each system via MCMC, our code \textsc{yMCMC} reads in a file containing the following four columns of measured values for a suite of emission lines: (1) the flux ratio, (2) the flux ratio uncertainty, (3) the equivalent width of the line in units of \AA, and (4) the uncertainty of the equivalent width of the line (\AA). The flux ratios and their corresponding errors are given relative to H$\beta$ for all optical emission lines, while P$\gamma$ is used for the NIR \HeI~$\lambda$10830 line. Since the input NIR flux ratio is not given relative to H$\beta$, our model separately calculates the predicted flux of \HeI~$\lambda$10830 and P$\gamma$ relative to H$\beta$, and combines these two predicted values to match the input format, $F$(\HeI~$\lambda$10830)~/~$F$(P$\gamma$):
\begin{equation}
    \frac{F(\textnormal{\HeI}~\lambda10830)}{F(P\gamma)}\,=\,\frac{F(\textnormal{\HeI}~\lambda10830)}{F(H\beta)}\,\Big/\,\frac{F(P\gamma)}{F(H\beta)}
\label{equation:scale_to_Pg}
\end{equation}
where the right hand side of the equation can be calculated using Equations \ref{equation:fluxHe} and \ref{equation:fluxH}, for the numerator and denominator respectively.

Additionally, our model predicts a total flux ratio of the blended H8+\HeI~$\lambda$3889 lines, which differs from the deblending technique employed by \citeauthor{2010JCAP...05..003A} (\citeyear{2010JCAP...05..003A}; see their Equation 4.1). To do this, our model individually predicts an H8 and \HeI~$\lambda$3889 emission line flux ratio and sums the two for a blended flux:
\begin{equation*}
    \frac{F(\textnormal{H8+\HeI}~\lambda3889)}{F(H\beta)}\,=\,\frac{F(\rm H 8)}{F(\rm H\beta)}\,+\,\frac{F(\textnormal{\HeI}~\lambda3889)}{F(\rm H\beta)}
\end{equation*}


Finally, we note that our observations of the PHLEK sample used a dichroic at 5600~\AA\, (see Section \ref{optical_observations} for details), which means that the H$\alpha$ and H$\beta$ emission lines are detected on separate blue and red arms of LRIS. For these systems, we adapt the MCMC code to model the flux ratios relative to H$\alpha$ for all optical emission lines that are detected on the red side of LRIS, in a manner equivalent to that of Equation \ref{equation:scale_to_Pg} for the NIR emission lines. As standard, every emission line on the blue side of LRIS is modelled relative to H$\beta$. Because of the dichroic, we lose the H$\alpha$~/~H$\beta$ Balmer line ratio in our analysis, and this has a minor impact on our ability to solve for the parameters. To test how the loss of $F(\rm H\alpha$)~/~$F(\rm H\beta$) affects our results, we generated a synthetic spectrum with emission line fluxes mirroring the format of our LRIS observations, and tested our MCMC's ability to recover the input model parameters. We show the results of this test in Appendix \ref{mock_mcmc}. As expected, we do not constrain parameters that depend on the Balmer lines as tightly -- for example, the 1$\sigma$ errors on $c$(H$\beta$) double when we lose information on $F(\rm H\alpha$)~/~$F(\rm H\beta$). However, we find that even without $F(\rm H\alpha$)~/~$F(\rm H\beta$), our recovered parameters are within 1$\sigma$ of the input parameters, and the errors on $y^{+}$ increase by a factor of just 1.06$-$1.20.

\begin{figure*}[!ht]
\includegraphics[width=1.0\textwidth]{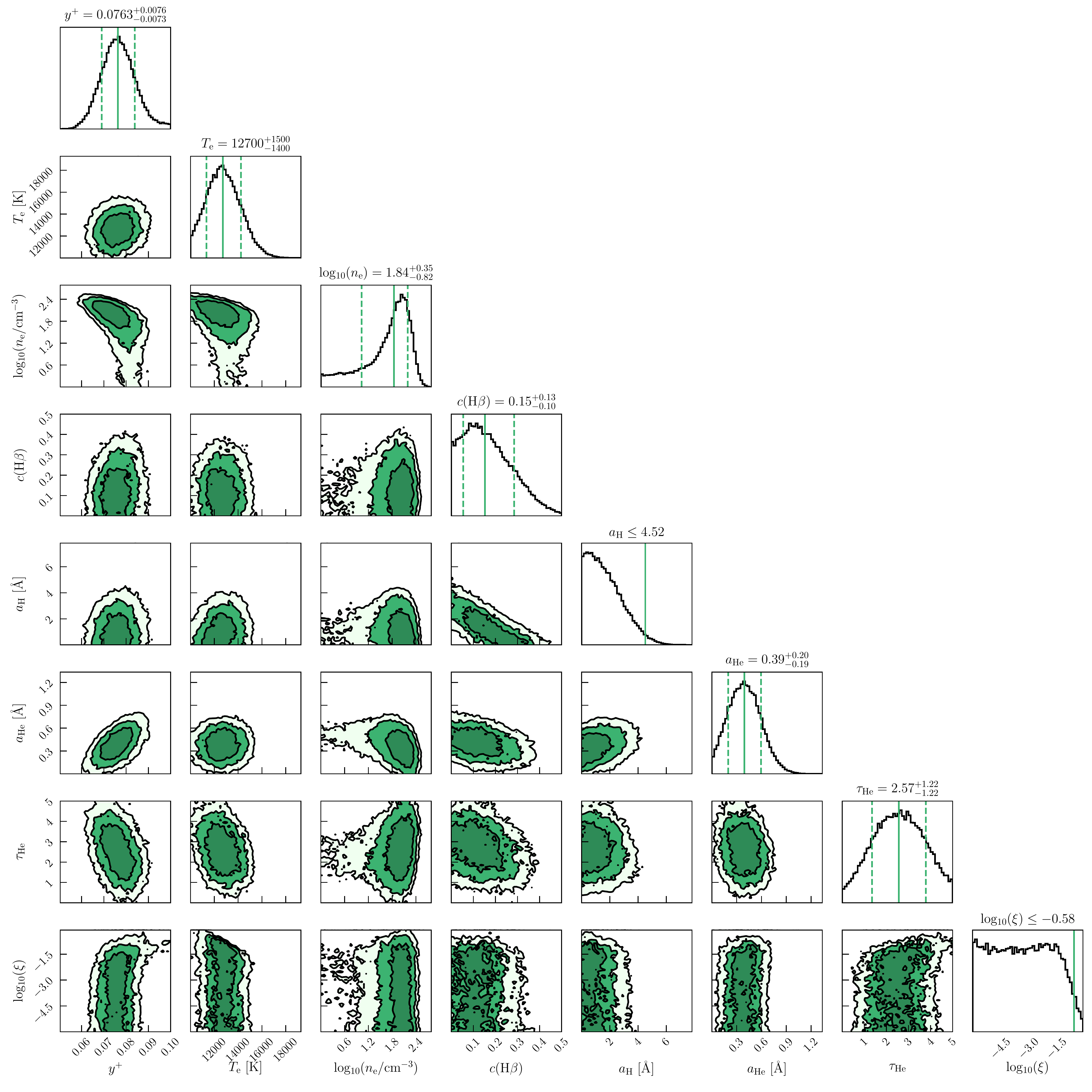}
\caption{Contours (off-diagonal panels) and histograms (diagonal panels) showing the best recovered model parameters of the galaxy J0118$+$3512. The contours show the 1$\sigma$, 2$\sigma$, and 3$\sigma$ levels. The solid green line in the histograms show the best recovered parameter value, and the dotted green lines show the $\pm1\sigma$ values. In the panels showing the results for the $a_{\rm H}$ and log$_{10}(\xi)$ parameters, the solid vertical line represents a $2\sigma$ upper limit. Observations of this galaxy include NIR data, which delivers a well-constrained value for the density parameter, log$_{10}(n_{\rm e}/\rm cm^{-3}$).}
\label{fig:J0118_parameter_contours}
\end{figure*}

Our MCMC analysis uses 500 walkers and 1000 steps to determine the best fit model parameters of each system. We take our burn-in to be a conservative 0.8$\times\,n_{\rm steps}\,=\,$ 800 steps and dispose of all samples before the burn-in, leaving us with $10^{5}$ samples. To ensure that our MCMC chains have converged, we require the best recovered parameters from the two halves of the $10^{5}$ samples to agree to within a few percent. In this exercise, all best recovered $y^{+}$ values agree to within half a per cent. We show an example contour plot and histogram of the recovered parameters of J0118$+$3512 in Figure \ref{fig:J0118_parameter_contours}.

\section{The Primordial Helium Abundance}
\label{analysis}
In the following sections, we describe the sample definition, the calculation of the metal abundances, and our determination of the primordial helium abundance.

\subsection{Qualification}
\label{qualification}
To identify the systems that are most suitable for determining the primordial helium abundance, we first require that all systems have a measured $EW(\rm H\beta)\geq50~\AA$. This ensures that systems have higher emission line flux-to-continuum level ratios, thus weak emission lines are less affected by underlying absorption. In particular, this minimizes the effect of underlying \HeI\ absorption and ensures that the measured \HeI\ emission line ratios do not under predict the true helium abundance \citep{2004ApJ...602..200I, 2007ApJ...662...15I}. This cut eliminates 3 systems from our PHLEK sample, 463 systems from the SDSS sample, and 4 from the HeBCD sample.

We also require that the recovered best fit parameters from the MCMC analysis are physical. Specifically, we remove all systems with recovered optical depths $\tau_{\rm He}\,>\,$4 and neutral to ionized hydrogen fractions $\xi\,>\,0.01$, if the 1$\sigma$ lower bound on the recovered value of $\xi$ does not encompass $\xi\,=\,$0.001. These specific values follow the work most recently highlighted by \cite{2015JCAP...07..011A}, although we have opted to completely eliminate systems with recovered parameters in the regimes stated above, while \cite{2015JCAP...07..011A} consider some of these systems as part of their flagged data set. We then assert that the MCMC analysis recovers parameters that are able to successfully reproduce all measured emission line ratios, according to the criteria listed in Sections \ref{sample1} and \ref{sample2} below.

\subsubsection{Sample 1}
\label{sample1}
\begin{figure*}[!ht]
\includegraphics[width=1.0\textwidth]{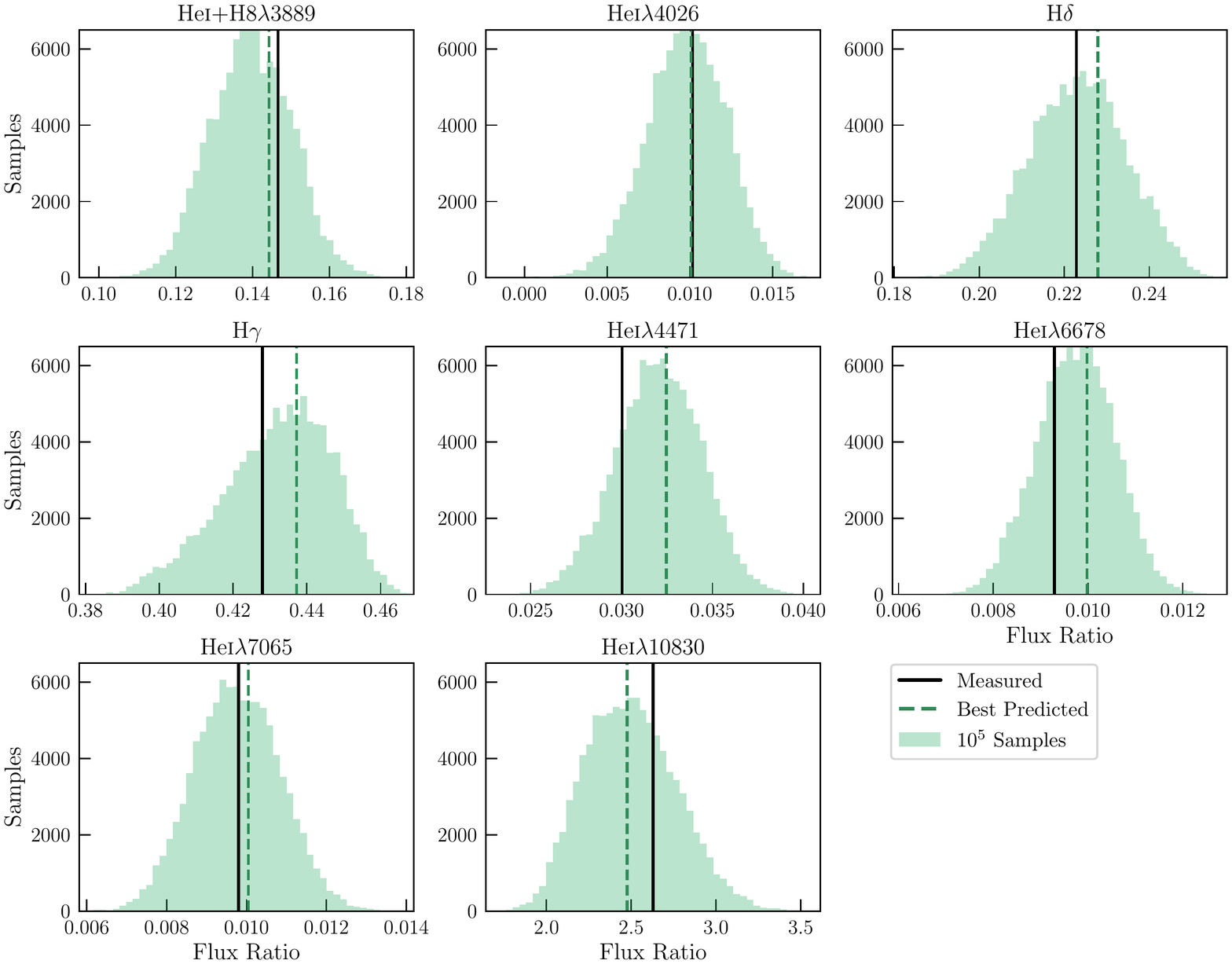}
\caption{Histograms showing the distributions of emission line flux ratios of the galaxy J0118$+$3512, derived from the final $10^{5}$ samples of our MCMC analysis. The green dashed lines show the value of the best recovered flux ratios and the solid black lines show the measured emission line flux ratio. H$\beta$ and H$\alpha$ do not appear in this figure since our LRIS blue and red side emission lines are measured relative to those two emission lines, and therefore those emission lines do not carry any information. Objects that qualify in Sample 1 require that the emission line ratios reproduced by our model are within 2$\sigma$ of the measured value. Note that the distributions shown by the histograms reflect the measurement errors.}
\label{fig:J0118_1e5_chains_measured_and_recovered}
\end{figure*}

Our most stringent criteria is that all of the measured emission line ratios are reproduced to within 2$\sigma$, given the best recovered parameters from the MCMC. This qualification criteria is demonstrated for the galaxy J0118$+$3512 in Figure \ref{fig:J0118_1e5_chains_measured_and_recovered}. We label the systems that qualify via these conditions ``Sample 1''. Sample 1 contains 3 galaxies from the PHLEK sample, 38 galaxies from the SDSS sample, and 13 galaxies from the HeBCD sample, resulting in a total of 54 systems. These systems and their best-recovered parameters are listed in part in Table \ref{table:s1_mcmc_recovered_params} and are available in full online. The full MCMC chains for Sample 1 are available on GitHub as part of the HCPB20 branch of our primordial helium code, \textsc{yMCMC}.


\subsubsection{Sample 2}
\label{sample2}
We also consider a more lenient qualification criteria, which requires that all emission line ratios are reproduced to within 2$\sigma$, with the exception of one emission line ratio, which must be reproduced to within 3$\sigma$. We call this ``Sample 2''.

Sample 2 consists of all of the systems in Sample 1, plus an additional 4 galaxies from our PHLEK sample, 48 galaxies from the SDSS sample, and 13 galaxies from the HeBCD sample. Thus, Sample 2 contains a total of 119 galaxies. These systems and their best recovered parameters from the MCMC are available online as part of Table \ref{table:s1_mcmc_recovered_params}. The full MCMC chains for Sample 2 are available on GitHub as part of the HCPB20 branch of \textsc{yMCMC}.

\begin{deluxetable*}{ccccccccc}
\tablewidth{12pt}
\tablecaption{Best recovered parameters from MCMC analysis} 
\tablehead{ 
\colhead{Galaxy} & \colhead{$y^{+}$} & \colhead{$T_{\rm e}$} & \colhead{log$_{10}(n_{\rm e}/\rm cm^{-3}$)} & \colhead{$c$(H$\beta$)} & \colhead{$a_{\rm H}$} & \colhead{$a_{\rm He}$} & \colhead{$\tau_{\rm He}$} & \colhead{log$_{10}(\xi$)} \\
\colhead{} & \colhead{} & \colhead{[K]} & \colhead{} & \colhead{} & \colhead{[\AA]} & \colhead{[\AA]} & \colhead{} & \colhead{}
}
\startdata
J0118$+$3512 & $0.0763^{+0.0076}_{-0.0073}$ & $12700^{+1500}_{-1400}$ & $1.84^{+0.35}_{-0.82}$ & $0.153^{+0.132}_{-0.099}$ & $1.5^{+1.5}_{-1.0}$ & $0.39^{+0.20}_{-0.19}$ & $2.6^{+1.2}_{-1.2}$ & $-3.3^{+1.8}_{-1.9}$ \\
J2030$-$1343 & $0.0725^{+0.0058}_{-0.0050}$ & $12400^{+1400}_{-1200}$ & $1.50^{+0.89}_{-1.01}$ & $0.280^{+0.048}_{-0.061}$ & $0.53^{+0.70}_{-0.38}$ & $0.19^{+0.16}_{-0.12}$ & $0.55^{+0.65}_{-0.38}$ & $-2.9^{+1.7}_{-2.1}$ \\
KJ29 & $0.081^{+0.011}_{-0.015}$ & $11440^{+1320}_{-940}$ & $1.76^{+0.54}_{-1.01}$ & $0.096^{+0.106}_{-0.069}$ & $0.34^{+0.44}_{-0.24}$ & $0.80^{+0.28}_{-0.34}$ & $1.6^{+1.8}_{-1.1}$ & $-3.1^{+1.9}_{-2.0}$ \\
spec-0301-51942-0531 & $0.0915^{+0.0059}_{-0.0064}$ & $13700^{+1500}_{-1500}$ & $2.19^{+0.61}_{-1.42}$ & $0.179^{+0.069}_{-0.070}$ & $0.58^{+0.92}_{-0.43}$ & $0.22^{+0.24}_{-0.16}$ & $2.1^{+1.2}_{-1.1}$ & $-1.60^{+0.71}_{-2.42}$ \\
spec-0364-52000-0187 & $0.0863^{+0.0045}_{-0.0029}$ & $11790^{+1110}_{-930}$ & $1.25^{+0.86}_{-0.85}$ & $0.378^{+0.032}_{-0.047}$ & $0.45^{+0.49}_{-0.32}$ & $0.250^{+0.103}_{-0.090}$ & $0.82^{+0.45}_{-0.43}$ & $-2.3^{+1.5}_{-2.6}$ \\
spec-0375-52140-0118 & $0.0842^{+0.0051}_{-0.0056}$ & $14400^{+1600}_{-1600}$ & $1.89^{+0.76}_{-1.20}$ & $0.208^{+0.033}_{-0.039}$ & $0.53^{+0.75}_{-0.39}$ & $0.51^{+0.26}_{-0.25}$ & $1.84^{+0.90}_{-0.93}$ & $-3.5^{+1.4}_{-1.6}$ \\
I Zw 18 SE1 & $0.0763^{+0.0031}_{-0.0028}$ & $17900^{+1900}_{-2000}$ & $1.82^{+0.14}_{-0.15}$ & $0.016^{+0.018}_{-0.011}$ & $3.65^{+0.54}_{-0.59}$ & $0.24^{+0.22}_{-0.16}$ & $0.64^{+0.55}_{-0.42}$ & $-4.71^{+0.92}_{-0.87}$ \\
SBS 0940+5442 & $0.0804^{+0.0033}_{-0.0023}$ & $17100^{+1400}_{-1400}$ & $1.937^{+0.090}_{-0.095}$ & $0.053^{+0.025}_{-0.028}$ & $2.11^{+0.98}_{-0.93}$ & $0.39^{+0.15}_{-0.14}$ & $0.29^{+0.30}_{-0.21}$ & $-3.75^{+0.94}_{-1.55}$ \\
Mrk 209 & $0.0820^{+0.0025}_{-0.0023}$ & $17400^{+1900}_{-1900}$ & $1.85^{+0.14}_{-0.15}$ & $0.018^{+0.018}_{-0.012}$ & $1.93^{+0.79}_{-0.84}$ & $0.26^{+0.12}_{-0.12}$ & $1.26^{+1.02}_{-0.81}$ & $-4.69^{+1.02}_{-0.90}$ \\
\enddata
\tablecomments{The best recovered values of the eight parameters sampled with our MCMC analysis. These parameters describe a subset of galaxies from Sample 1, defined to be systems where the best recovered parameters can reproduce all the measured emission line flux ratios to within 2$\sigma$. This table lists three systems from our PHLEK sample (top three rows; see Section \ref{oursample}), the SDSS sample (middle three rows; see Section \ref{sdss}), and the HeBCD sample (bottom three rows; see Section \ref{hebcd}).} 
\label{table:s1_mcmc_recovered_params}
\end{deluxetable*}

\subsection{Abundance Measurements}
To calculate ionic abundances, we utilize the emission line analysis package \textsc{PyNeb} \citep{2015A&A...573A..42L}.\footnote{\textsc{PyNeb} can be downloaded from: \url{http://www.iac.es/proyecto/PyNeb/}} To obtain a value and error on an ionic abundance, we calculate $10^{5}$ Monte Carlo realizations of each abundance by perturbing the measured flux ratios by their errors. For each realization, we use relevant parameters derived from our MCMC samples, namely the electron density, $n_{\rm e}$, and reddening parameter, $c$(H$\beta$). Our reported abundances and their errors are the mean and standard deviation of the $10^{5}$ Monte Carlo realizations.

\HII\ regions are expected to be in the low-density regime, where density diagnostics observed at optical wavelengths, such as the [\SII]~$\lambda\lambda$6717,\,6731 doublet, are not very sensitive to $n_{\rm e}$ (see Figure 5.3 of \citealt{1989agna.book.....O}). As such, the $n_{\rm e}$ value recovered by the MCMC analysis is loosely constrained when our observations do not include lines that are strongly sensitive to $n_{\rm e}$ in the low density regime. Previous works in the literature usually assume $n_{\rm e}\,=\,100$ for ionic abundance calculations instead of the measured electron density, an assumption that is within the 1$\sigma$ bounds of their measured values. This choice is also within the range of densities expected of \HII\ regions, $n_{\rm e}\,=\,100$~--~$10,000$~cm$^{-3}$ \citep{1989agna.book.....O}. However,  $n_{\rm e}$ can be pinned down when the density-sensitive \HeI~$\lambda$10830 line is included in the analysis (see our distribution and best recovered value of log($n_{\rm e}/\rm cm^{-3}$) in Figure \ref{fig:J0118_parameter_contours} as an example, as well as the results of our trial MCMC runs on mock data including the \HeI~$\lambda$10830 line in Appendix \ref{mock_mcmc}). We therefore adopt the $n_{\rm e}$ values sampled by our MCMC as input to \textsc{PyNeb} for the determination of the ionic abundances. 

\subsubsection{Oxygen}
\label{oxygen_abundance}
The total oxygen abundance $\rm O/\rm H$ is the sum of the singly and doubly ionized ionic abundances:
\begin{equation}
    \frac{\rm O}{\rm H}\,=\,\frac{\rm O^{+}}{\rm H^{+}}\,+\,\frac{\rm O^{++}}{\rm H^{+}}
\label{equation:oxygen_abundance}
\end{equation}

The values of $\rm O^{+}/\rm H^{+}$ and $\rm O^{++}/\rm H^{+}$ (hereafter abbreviated as $\rm O^{+}$ and $\rm O^{++}$) of each galaxy depend on the measured emission line flux ratios relative to $\rm H\beta$, the electron temperature, and the electron density. We adopt a two-zone approximation of an \HII\ region, with two distinct electron temperatures characterizing the high- and low-ionization zones. The $\rm O^{++}$ abundance is calculated using the [\OIII]~$\lambda\lambda$4959,\,5007 flux ratios in combination with the high ionization zone temperature, $t_{3}$, where we calculate values of $t_{3}$ using the temperature sensitive [\OIII]~$\lambda$4363 line. Thus, the value of $t_{3}$ differs from the electron temperature parameter in our MCMC model, $T_{\rm e}$, but the difference is expected to be small.

As mentioned previously, we calculate $10^{5}$ values of the $\rm O^{++}$ abundance, each time adopting an electron density value as sampled in the $10^{5}$ density realizations of the MCMC chain. The measured [\OIII] flux ratios are perturbed each time by drawing a new value from a Gaussian distribution with a mean of the measured flux value and standard deviation of its measurement error. We also calculate a new value of $t_{3}$ at each step in the MCMC using the perturbed [\OIII] flux ratios.

The $\rm O^{+}$ abundance is calculated using the [\OII]~$\lambda\lambda$3727, 3729 doublet and the low-ionization zone temperature, $t_{2}$. A direct measure of $t_{2}$ requires a detection of the [\OII]~$\lambda\lambda$7320,\,7330\AA\, lines or the [\NII]~$\lambda$5755\AA\, line (used in conjunction with the [\NII]~$\lambda\lambda$6548,\,6584 doublet). Since we do not detect these lines, we infer $t_{2}$ from $t_{3}$ following the relation from \cite{1992MNRAS.255..325P}, which is based on the photoionization model grids by \cite{1990A&AS...83..501S}:

\begin{equation}
    t_2\,=\,20,000~\rm K \Big/ \Big(\frac{10,000~\rm K}{\textit{t}_3}\,+\,0.8\Big)
\end{equation}
The total oxygen abundance of each system is calculated by summing the singly and doubly ionized oxygen abundances (i.e., Equation \ref{equation:oxygen_abundance}). The final reported oxygen abundance and its corresponding error, is calculated by taking the mean and standard deviation of the $10^{5}$ Monte Carlo realizations. 
All abundance calculations are made using \textsc{PyNeb}'s \texttt{getIonAbundance()} method. We report the ionic and total oxygen abundances of a subset of our systems in Table \ref{table:s1_abunds_metallicity} and in full online.

\subsubsection{Helium}
\label{helium_abundance}
The total helium abundance, $y$, is the sum of the abundances of singly ionized helium $y^{+}$ and doubly ionized helium $y^{++}$ (see Equation \ref{equation:helium_abundance}). $y^{+}$ is recovered as a parameter of the MCMC analysis, and the presence of $y^{++}$ in an \HII\ region can be inferred via emission at \HeII~$\lambda$4686\AA\, \citep{1992MNRAS.255..325P, 2013AJ....146....3S}. Therefore, if the \HeII~$\lambda$4686 line is detected, we calculate and include the $y^{++}$ abundance in the total helium abundance. A non-detection of \HeII~$\lambda$4686\, in the spectrum is assumed to indicate a negligible $y^{++}$ abundance.

As with calculating the oxygen abundance, we assume an electron density $n_{\rm e}$ as recovered by the MCMC. However, for the helium abundances, we also assume the electron temperature $T_{\rm e}$ from the MCMC chains. We make $10^{5}$ realizations of the $y^{++}$ abundance by perturbing the measured \HeII\, flux ratios by the error in its measurement. While we expect doubly ionized helium to occupy a region of higher temperatures than $T_{\rm e}$ (i.e., the temperature at which singly ionized helium is found), this assumption has a negligible effect on the total helium abundance, since the $y^{++}$ abundance typically contributes a $\sim1$ per cent correction to the overall helium abundance.

Furthermore, some of the helium in \HII\ regions may be in the neutral state; thus, the total helium abundance may require a correctional factor for undetected neutral helium. To assess whether a neutral helium component is present, we follow the use of the radiation softness parameter, $\eta$ \citep{1988MNRAS.231..257V}, defined as:

\begin{equation}
    \ \eta\,=\,\frac{O^{+}}{S^{+}}\times\frac{S^{++}}{O^{++}}
\end{equation}

\noindent to estimate the hardness of the ionizing radiation. The $S^{++}$ abundance depends on the temperature of the gas, $t_{S^{++}}$. A direct measure of $t_{S^{++}}$ requires the detection of [\SIII] emission at $\lambda$6312\AA, $\lambda$9069\AA, and $\lambda$9532\AA, the latter two of which fall outside the wavelength coverage of our instrument setup.\footnote{In some cases, we detect [\SIII]$\lambda$9532 when we have NIR observations, but we have no coverage of [\SIII]$\lambda$9069.} The $S^{++}$ abundance is extremely sensitive to temperature \citep{1992AJ....103.1330G}, therefore, rather than assuming the value of $t_{S^{++}}$ to be $t_{3}$ or $t_{2}$, it is necessary to estimate the temperature of the $S^{++}$ zone following the relation from \cite{1992AJ....103.1330G}:

\begin{equation*}
    t_{S^{++}}\,=\,0.83t_3\,+\,0.17
\end{equation*}

\noindent We assume $t_{S^{+}}\,=\,t_{S^{++}}$, following the expected ionization structure in a two-zone photoionization model (see e.g., Figure 2 of \citealt{1992AJ....103.1330G}). 
Adopting this temperature, the $S^{++}$ and $S^{+}$ abundances can be calculated with the [\SIII]~$\lambda$6312 and [\SII]~$\lambda\lambda$6717,\,6731 emission line fluxes.


Based on photoionization models, \cite{1992MNRAS.255..325P} concluded $\eta$ to be suitable for determining whether a correctional factor is necessary for undetected neutral helium; if log($\eta$)\,$<$\,0.9, the neutral helium abundance can be assumed to be negligible (see Figure 6 of \citealt{1992MNRAS.255..325P}). We choose to exclude the systems from our sample that were found to have a non-negligible neutral helium abundance following this metric (4 total systems, all of which are from the SDSS sample), due to the additional uncertainties introduced when assuming a correctional factor.

The ionic and total helium abundances of our systems are partially listed in Table \ref{table:s1_abunds_metallicity} and are available in full online.

\begin{deluxetable*}{ccccccc}
\tablewidth{12pt}
\tablecaption{Ionic and total abundances of oxygen and helium} 
\tablehead{ 
\colhead{Galaxy} & \colhead{O$^{+}$~/~H$^{+}$} & \colhead{O$^{++}$~/~H$^{+}$} & \colhead{$(\rm O/\rm H)$} & \colhead{$y^{+}$} & \colhead{$y^{++}$} & \colhead{$y$} \\
\colhead{} & \colhead{(~$\times10^{5}$~)} & \colhead{(~$\times10^{5}$~)} & \colhead{(~$\times10^{5}$~)} & \colhead{} & \colhead{} & \colhead{}
}
\startdata
J0118$+$3512 & $1.44^{+0.38}_{-0.27}$ & $5.1^{+2.2}_{-1.3}$ & $6.5^{+2.2}_{-1.3}$ & $0.0764^{+0.0076}_{-0.0075}$ & 0.0028$^{+0.0002}_{-0.0002}$ & $0.0792^{+0.0076}_{-0.0075}$ \\
J2030$-$1343 & $2.28^{+0.55}_{-0.41}$ & $7.9^{+3.1}_{-2.0}$ & $10.2^{+3.1}_{-2.1}$ & $0.0725^{+0.0058}_{-0.0050}$ & 0.0018$^{+0.0001}_{-0.0001}$ & $0.0743^{+0.0058}_{-0.0050}$ \\
KJ29 & $2.10^{+0.43}_{-0.41}$ & $5.5^{+1.7}_{-1.5}$ & $7.6^{+1.8}_{-1.6}$ & $0.081^{+0.011}_{-0.015}$ & \nodata & $0.081^{+0.011}_{-0.015}$ \\
spec-0301-51942-0531 & $2.34^{+0.51}_{-0.33}$ & $7.1^{+2.6}_{-1.7}$ & $9.5^{+2.7}_{-1.7}$ & $0.0915^{+0.0059}_{-0.0064}$ & 0.0014$^{+0.0003}_{-0.0003}$ & $0.0929^{+0.0059}_{-0.0064}$ \\
spec-0364-52000-0187 & $2.73^{+0.51}_{-0.41}$ & $12.4^{+3.7}_{-2.9}$ & $15.1^{+3.7}_{-2.9}$ & $0.0863^{+0.0045}_{-0.0029}$ & 0.0007$^{+0.0001}_{-0.0001}$ & $0.0870^{+0.0045}_{-0.0030}$ \\
spec-0375-52140-0118 & $1.24^{+0.29}_{-0.19}$ & $7.7^{+2.9}_{-1.8}$ & $8.9^{+3.0}_{-1.8}$ & $0.0842^{+0.0051}_{-0.0056}$ & 0.0008$^{+0.0002}_{-0.0002}$ & $0.0850^{+0.0051}_{-0.0056}$ \\
I Zw18 SE1 & $0.465^{+0.083}_{-0.055}$ & $1.31^{+0.39}_{-0.27}$ & $1.78^{+0.40}_{-0.28}$ & $0.0763^{+0.0031}_{-0.0028}$ & 0.0008$^{+0.0002}_{-0.0002}$ & $0.0772^{+0.0031}_{-0.0028}$ \\
SBS 0940+5442 & $0.436^{+0.058}_{-0.045}$ & $3.37^{+0.71}_{-0.53}$ & $3.81^{+0.71}_{-0.53}$ & $0.0804^{+0.0033}_{-0.0023}$ & 0.0005$^{+0.0001}_{-0.0001}$ & $0.0810^{+0.0033}_{-0.0023}$ \\
Mrk 209 & $0.679^{+0.123}_{-0.087}$ & $4.38^{+1.30}_{-0.93}$ & $5.06^{+1.30}_{-0.94}$ & $0.0820^{+0.0025}_{-0.0023}$ & 0.0011$^{+0.0000}_{-0.0000}$ & $0.0831^{+0.0025}_{-0.0023}$ \\
\enddata
\tablecomments{The singly and doubly ionized oxygen abundances, total oxygen abundance, singly and doubly ionized helium abundances, and total helium abundance for a subset of galaxies from Sample 1. This table lists three systems from our PHLEK sample (top three rows; see Section \ref{oursample}), the SDSS sample (middle three rows; see Section \ref{sdss}), and the HeBCD sample (bottom three rows; see Section \ref{hebcd}). The online version of this table also contains the remaining galaxies in Sample 1 as well as Sample 2.} 
\label{table:s1_abunds_metallicity}
\end{deluxetable*}


\subsection{Extrapolation to $y_{\rm P}$}
\label{extrap_to_yp}
The standard approach for determining the primordial helium abundance is to perform a linear regression to a set of measured oxygen and helium abundances. This technique was initially proposed by \cite{1974ApJ...193..327P, 1976ApJ...203..581P} and is still used by the most recent primordial helium abundance investigations. The analysis follows the expectation from BBN calculations that most of the helium in the Universe is produced during BBN, while essentially no oxygen is produced. Through the chemical evolution of stars, there is a net production of \fourHe, but this contribution is relatively minor compared to the quantity of \fourHe\ produced during BBN. Therefore, the post-BBN contribution to the \fourHe\ abundance can be modelled as a small (linear) deviation from the BBN value that increases with increasing metallicity. 
We also note that \cite{2018MNRAS.478.5301F} have recently proposed that a tighter relation exists between the helium abundance and the sulphur abundance. Their work suggests that, as far as chemical evolution is concerned, sulphur may trace helium better than oxygen. However, in our work, we do not have access to the emission lines required to measure the sulphur abundance, and we therefore use the $\rm O/\rm H$ abundance in what follows.

Our determination of the primordial helium abundance, $y_{\rm P}$, is based on a fit to the measured $\rm O/\rm H$ and ${\rm He/\rm H\,\equiv}\,y$ number abundance ratios of the galaxies that qualify for Sample 1 and 2. We note that our choice to use the helium number abundance ratio differs from the typical format historically found in the literature, where the primordial helium abundance is expressed as the primordial helium mass fraction, $Y_{\rm P}$. For reference, the helium mass fraction, $Y$, can be converted from $y$ using:
\begin{equation*}
\ Y = \frac{4y\,(1-Z)}{1+4y}
\end{equation*}
and
\begin{equation*}
\ Z = c \times (\rm O/\rm H)
\end{equation*}
\noindent Here, $Z$ is the metallicity dependent heavy element mass fraction, which is linearly proportional to the constant $c$, which depends on chemical evolution (see directly below for a further discussion of this constant).

We have decided to use the helium number abundance $y$ instead of the helium mass fraction $Y$, as done historically, for the following reasons:
\begin{itemize}
    \item Observations of the helium abundance are intrinsically measuring a number abundance ratio.
    \item Calculations of the helium abundance are computed as a ratio of volume densities ($n_{^{4}\rm He}/n_{^{1}\rm H}$), and later converted to a mass fraction to match the observationally reported mass fractions (see e.g., \citealt{2018PhR...754....1P}).
    \item The primordial helium mass fraction ($Y_{\rm P}$) is not actually the fraction of mass in the form of \fourHe. It is defined as the ratio of volume densities $Y_{\rm P}\,=\,4\,n(^{4}{\rm He})/n_{\rm b}$, where $n_{\rm b}$ is the baryon density. Thus, the term ``mass fraction'' is a misnomer that should probably be avoided as we enter the era of precision cosmology.
    \item Our choice eliminates the dependence on $Z$, whose value has varied across primordial helium works. For reference, \cite{1992MNRAS.255..325P} and \cite{2015JCAP...07..011A} both take $c\,=\,20$ for $Z\,=\,20\,\times\,($\rm O\,/\,\rm H$)$, \cite{2007ApJ...662...15I} adopts $c\,=\,$18.2, and \cite{2013A&A...558A..57I} allow for a $c$ value that linearly scales with the metallicity, $c\,=8.64\times12\,+\,\rm log_{10}($\rm O\,/\,\rm H$)\,-$\,47.44.
\end{itemize}
For these reasons, we have therefore chosen to quote our primordial helium abundance in the form we most directly measure and most appropriate to compare to theoretical values --- the primordial helium number abundance ratio, $y_{\rm P}$. However, a comparison of our measured $y_{\rm P}$ to previously reported values of $Y_{\rm P}$ can be simply calculated with the following equation:
\begin{equation}
\label{equation:y_to_Y}
\ Y_{\rm P} = \frac{4y_{\rm P}}{1+4y_{\rm P}}
\end{equation}

Our linear fits to the two galaxy samples described in Sections \ref{sample1} (Sample 1) and \ref{sample2} (Sample 2) are optimized using \textsc{emcee} given the likelihood function of our linear model:
\begin{equation}
\label{equation:log_likelihood_yp}
\textnormal{log}(\mathcal{L})\,=\,-\frac{1}{2}\sum_{n}\Big[\frac{(y_{n} - mx_{n} - b)^{2}}{\sigma_{y_{n}}^{2}+\sigma_{\rm intr}^{2}} - \textnormal{log}(\sigma_{y_{n}}^{2}+\sigma_{\rm intr}^{2})\Big]
\end{equation}
\noindent Here, the summation is over all individual galaxies in each sample. Our linear model is given by $mx_{n} + b$, where the $x_{n}$ are our measured $\rm O/\rm H$ values, the slope $m\,\equiv\,$d$y/ \rm d(\rm O/\rm H)$, and the intercept $b\,\equiv\,y_{\rm P}$. We capture the error on our calculated $\rm O/\rm H$ abundances by drawing new values of $\rm O/\rm H$ from a Gaussian with a mean of the calculated values and standard deviation of the calculated errors during each step of the MCMC procedure. The total measured error of $y_{n}$ is captured by the term $\sigma_{y_{n}}$. We also introduce the term $\sigma_{\rm intr}$ to our likelihood function to quantify the intrinsic scatter of our sample of $y$ measurements to account for unknown systematic uncertainties that are introduced by our model, following the method presented in Section 4.3 of \cite{2018ApJ...855..102C}.

\begin{figure*}[!h]
\includegraphics[width=1.0\textwidth]{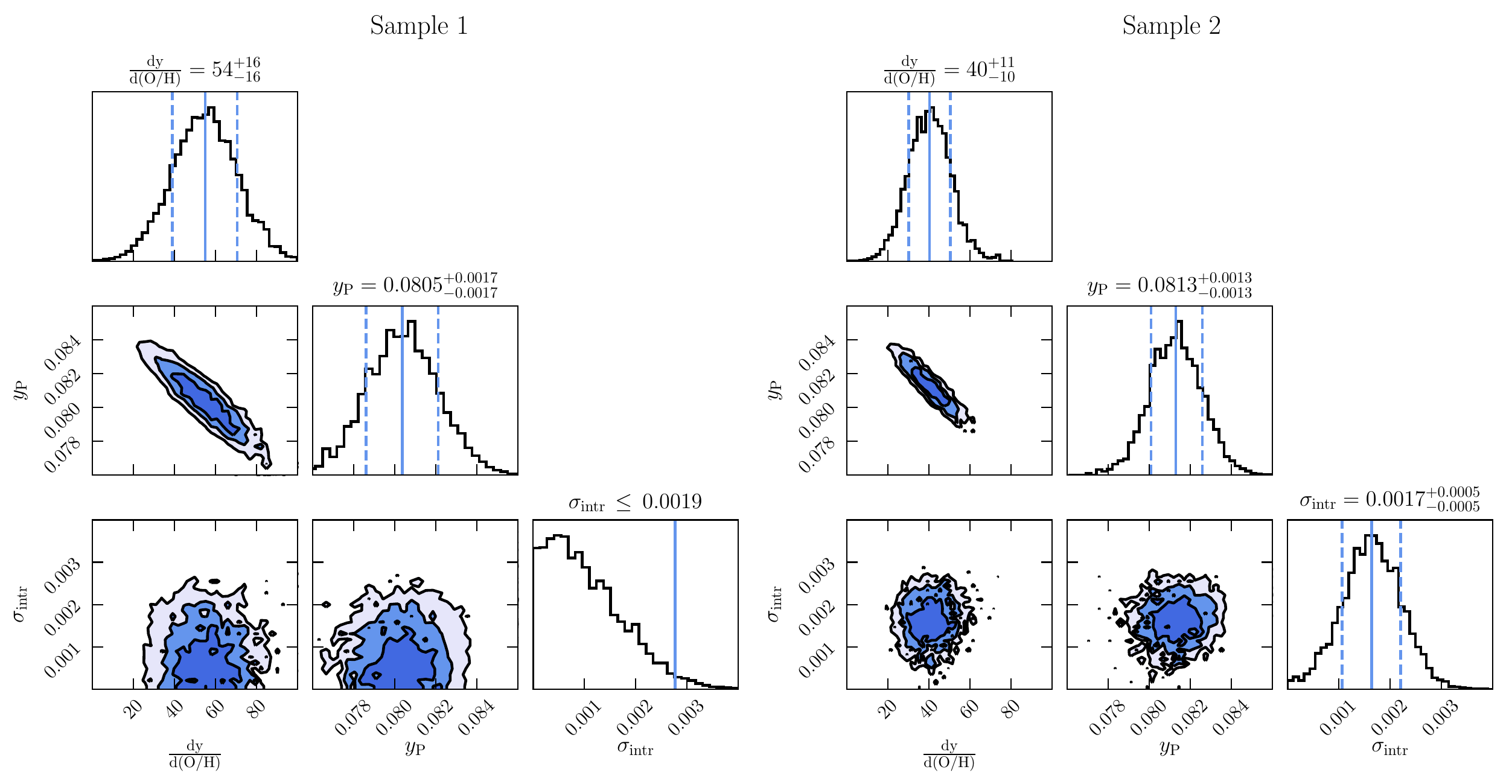}
\caption{Contours (off-diagonal panels) and posterior distributions (diagonal panels) showing the best fit slope (d$y$/d$(\rm O/\rm H)$), intercept ($y_{\rm P}$), and intrinsic scatter ($\sigma_{\rm intr}$), as recovered from the MCMC. The left and right panels show the MCMC results for Sample 1 and Sample 2, respectively, as defined in Sections \ref{sample1} and \ref{sample2}. For Sample 1, we report a 2$\sigma$ upper limit on $\sigma_{\rm intr}$ since it is consistent with zero.
The contours show the 1$\sigma$, 2$\sigma$, and 3$\sigma$ levels. The solid vertical blue lines in the diagonal panels indicate the best recovered values, while the dotted blue lines represent the $\pm1\sigma$ values on the parameters. The linear model described by these parameters (given in Equation \ref{equation:log_likelihood_yp}) is overplotted in Figure \ref{fig:OH_vs_yp}.}
\label{fig:linear_fit_emcee}
\end{figure*}

To solve for the parameters that best describe our linear model and the intrinsic scatter, we use 1000 walkers each taking 1000 steps in the MCMC. We set the following uniform priors on the model parameters:
\begin{gather*} 
0 \leq \frac{\textnormal{d}y}{\rm d(O/H)} \leq 100 \\
0.06 \leq y_{\rm P} \leq 0.10 \\
0 \leq \sigma_{\rm intr} \leq 0.01
\end{gather*}
The range of allowed $y_{\rm P}$ values matches the range of $y^{+}$ values of our model described in Section \ref{model}. Similarly, we allow a generous range of possible d$y$/d($\rm O/\rm H$) values. The range in $\sigma_{\rm intr}$ is chosen to be comparable to the measurement error of the $y$ values, $\sigma_{y_{n}}$. We find a mean of $\langle\,\sigma_{y_{\rm n}}\,\rangle$\,=\,0.005, and allow for the range of $\sigma_{\rm intr}$ to be twice that value, although it is desirable for this parameter to be less than $\sigma_{y_{\rm n}}$. We conservatively use a burn-in of 800 steps. The distribution of the explored parameter space and the best recovered parameters using Sample 1 and Sample 2 are shown in Figure \ref{fig:linear_fit_emcee}.

\begin{figure*}[!h]
\includegraphics[width=1.0\textwidth]{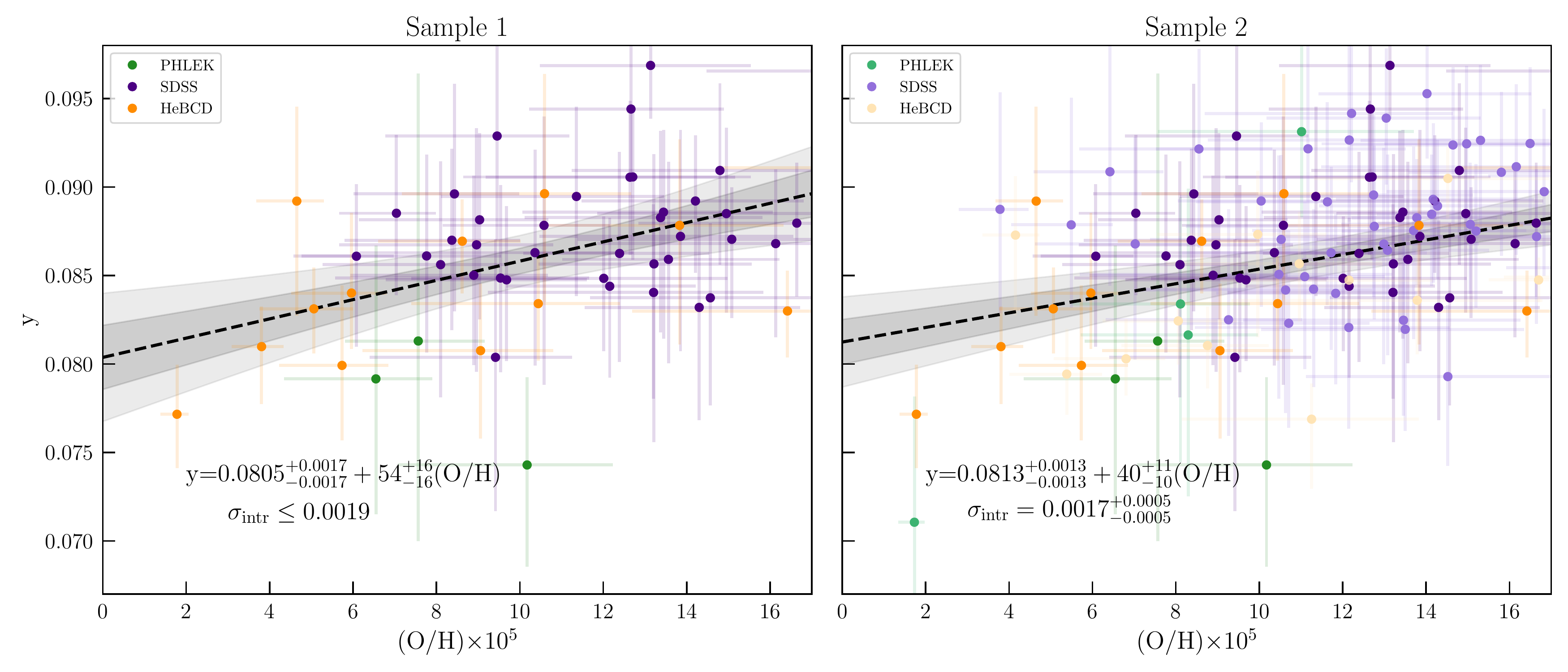}
\caption{Our extrapolation to the primordial helium abundance $y_{\rm P}$ using Sample 1 (left panel) and Sample 2 (right panel), which are described in Sections \ref{sample1} and \ref{sample2}, respectively. The green, purple, and orange, circles with error bars show our PHLEK sample, the SDSS sample, and the HeBCD sample of galaxies, respectively. The black dashed line indicates the best fit linear extrapolation to $y_{\rm p}$ while the surrounding shaded grey regions show the 1$\sigma$ and 2$\sigma$ errors on the linear fit. In the right panel, the darker points represent Sample 1, while the lighter points represent Sample 2. The expressions shown describe the best fit linear models along with the intrinsic scatter $\sigma_{\rm intr}$, which captures possible systematic uncertainties that are currently unaccounted for by our model.}
\label{fig:OH_vs_yp}
\end{figure*}

In Figure \ref{fig:OH_vs_yp}, we plot Samples 1 and 2, along with their best fit linear models and extrapolations to $y_{\rm P}$. The optimal parameter values recovered from the MCMC for Sample 1 are:
\begin{gather*}
    y_{\rm P}\,=\,0.0805^{+0.0017}_{-0.0017} \\
    \frac{\textnormal{d}y}{\rm d(O/H)}\,=\,54^{+16}_{-16} \\
    \sigma_{\rm intr}\,\leq\,0.0019\, (2\sigma\,\rm CL)
\end{gather*}
This model has a $\chi^{2}/\rm dof\,=\,0.77$. For Sample 2, we recover:
\begin{gather*}
    y_{\rm P}\,=\,0.0813^{+0.0013}_{-0.0013} \\
    \frac{\textnormal{d}y}{\rm d(O/H)}\,=\,40^{+11}_{-10} \\
    \sigma_{\rm intr}\,=\,0.0017^{+0.0005}_{-0.0005}
\end{gather*}
with $\chi^{2}/\rm dof\,=\,0.82$. While the linear fits to Sample 1 and Sample 2 are comparable, we adopt $y_{\rm P}$ from Sample 1 as our reported value and in all further analyses. This choice is motivated by our model being able to more confidently reproduce all of the observed emission line fluxes of Sample 1. This increases our confidence in the recovered parameters, including the primordial helium abundance. This confidence is also reflected in the recovered value of $\sigma_{\rm intr}$, which is consistent with zero for Sample 1 but non-zero for Sample 2. The recovered $y_{\rm P}$ values from Sample 1 and Sample 2 are within 1$\sigma$ of each other, but we note that Sample 1 and Sample 2 are not independent of one another (i.e., Sample 1 is a subset of Sample 2).

\subsection{Comparison to Existing Measurements of $Y_{\rm P}$}
\label{comparison_to_lit}
We now compare our result to existing primordial helium abundance measurements that are reported in the literature. To allow for a comparison of the primordial helium number abundance ratio, $y_{\rm P}$, we convert all literature measurements of $Y_{\rm P}$ to $y_{\rm P}$ using Equation \ref{equation:y_to_Y}. The literature results are summarized in Table \ref{table:Yp_from_literature}. Our result agrees with measurements derived from emission line observations of \HII\ regions in nearby galaxies \citep{2015JCAP...07..011A, 2016RMxAA..52..419P, 2019MNRAS.487.3221F, 2019ApJ...876...98V}, absorption line observations of a near-pristine gas cloud along the line-of-sight to a background quasar \citep{2018NatAs...2..957C}, the primordial helium abundance derived from the damping tail of the CMB recorded by the \textit{Planck} satellite \citep{2016AA...594A..13P}, and SBBN calculations of the primordial abundances \citep{2016RvMP...88a5004C, 2018PhR...754....1P} that assume a baryon-to-photon ratio $\eta\,=\,(5.931\,\pm\,0.051)\,\times\,10^{-10}$, which is based on the observationally measured abundance of primordial deuterium, $(\rm D/\rm H)_{\rm P}\,=\,(2.527\,\pm\,0.030)\,\times\,10^{-5}$ \citep{2018ApJ...855..102C}.

\begin{deluxetable*}{cccc}[ht!]
\tablecaption{Primordial helium abundance results reported in the literature} 
\tablehead{ 
\colhead{$y_{\rm P}$} & \colhead{Observation/Method} & \colhead{Number of Systems} & \colhead{Citation} \\
}
\startdata 
$0.0856\pm0.0010$ & \HII\ region & 28 & \citealt{2014MNRAS.445..778I} \\
$0.0811\pm0.0018$ & \HII\ region & 15 & \citealt{2015JCAP...07..011A} \\
$0.0809\pm0.0013$ & \HII\ region & 5 & \citealt{2016RMxAA..52..419P} \\
$0.0802\pm0.0022$ & \HII\ region & 18 & \citealt{2019MNRAS.487.3221F} \\
$0.0812\pm0.0011$ & \HII\ region in NGC 346 & 1 & \citealt{2019ApJ...876...98V} \\
$0.0805^{+0.0017}_{-0.0017}$ & \HII\ region & 54 & This work \\
$0.0793\pm0.011\,(2\sigma)$ & CMB & \nodata & \citealt{2018arXiv180706209P} \\
$0.085^{+0.015}_{-0.011}$ & Absorption line system & 1 & \citealt{2018NatAs...2..957C} \\
\hline
$0.0820\pm0.000074$ & SBBN calculation & \nodata & \citealt{2016RvMP...88a5004C} \\
$0.0820\pm0.000075$ & SBBN calculation & \nodata & \citealt{2018PhR...754....1P} \\
\enddata
\tablecomments{A summary of primordial helium abundance results reported in recent literature, the method by which the values are measured or calculated, and their reference. The \textit{Planck} measurement is the TT,TE,EE+lowE value from Equation 80a of \cite{2018arXiv180706209P} and is BBN-independent. All values are quoted with 1$\sigma$ confidence limit, except the CMB value, which is quoted with 2$\sigma$ confidence limit, as indicated.} 
\label{table:Yp_from_literature}
\end{deluxetable*}

The primordial helium abundance that we report here is in 2.6$\sigma$ disagreement with the \cite{2014MNRAS.445..778I} result, $y_{\rm P}\,=\,0.0856\pm0.0010$. We note that the HeBCD sample included in this work was compiled by \cite{2014MNRAS.445..778I} and was subsequently the sample analyzed by \cite{2015JCAP...07..011A}. \cite{2015JCAP...07..011A} also find a discrepancy with the \citeauthor{2014MNRAS.445..778I} results (2.2$\sigma$), and suggest several possible reasons for the disagreement. Given that our model closely follows that of \cite{2010JCAP...05..003A, 2012JCAP...04..004A, 2013JCAP...11..017A, 2015JCAP...07..011A}, we expect many of these reasons to equally be relevant in the comparison between our result and that of \cite{2014MNRAS.445..778I}. 
For example, \citeauthor{2014MNRAS.445..778I} first use the observed Balmer line ratios to solve for the amount of reddening and underlying hydrogen stellar absorption present, while assuming that the underlying absorption is the same for all hydrogen lines. After correcting observations for reddening, they then use Monte Carlo to find the best fit value of $y^{+}$, given $T_{\rm e}$, $n_{\rm e}$, and $\tau_{\rm He}$ and the observed \HeI\ lines. This differs from the MCMC method adopted in this work, which solves for all parameters using all observed emission lines simultaneously. Within their model, \citeauthor{2014MNRAS.445..778I} implement a correction for hydrogen emission resulting from collisional excitation based on \textsc{cloudy} photoionization modelling. There are also slight differences in the assumed coefficients for underlying stellar absorption between our models, the incorporation and scaling of the NIR lines to H$\beta$, and the calculation of $t_{2}$ and $t_{S^{++}}$ from $t_{3}$. Finally, \citeauthor{2014MNRAS.445..778I} apply a cut on their sample prior to solving for the best fit parameters, based on properties such as their measured $ EW$(H$\beta$) and ionization parameter. Our approach, on the other hand, solves for the best fit parameters of every galaxy and subsequently use this information to decide if a system qualifies. We refer readers to \cite{2006A&A...448..955I, 2013A&A...558A..57I, 2014MNRAS.445..778I} for details of their model and approach.

It is reassuring that our extrapolation to $y_{\rm P}$ is in agreement with numerous existing values reported in the literature. Most of these are based on distinct samples, a variety of sample sizes, and adopt different analysis methods. This does not, however, rule out the need for improvements in future primordial helium research; we discuss current model deficiencies that could warrant additional enhancements in Section \ref{future_improvements}.

\section{Discussion}
\label{discussion}
In this section, we use our determination of $y_{\rm P}$ to place a limit on physics beyond the Standard Model and discuss future improvements that can be made to push measurements of $y_{\rm P}$ to sub-percent level accuracy.

\subsection{Implications for the Standard Model -- BBN bounds on $\Omega_{\rm b} h^{2}$ and $N_{\rm eff}$}
\label{constraint_on_Omegab_Neff}
Physics beyond the Standard Model at the time of BBN can be identified by comparing observational measurements of the primordial abundances with the SBBN predicted values. The primordial element abundances produced during BBN are captured primarily by two parameters: the baryon density, $\Omega_{\rm b} h^{2}$, and the effective number of neutrino species, $N_{\rm eff}$. By adopting a measurement of $\Omega_{\rm b} h^{2}$ from the CMB \citep{2018arXiv180706209P} and assuming $N_{\rm eff}\,=\,3.046$ (i.e. the Standard Model value; \citealt{2002APh....17...87C, 2016RvMP...88a5004C, 2018PhR...754....1P}), BBN is a parameter free theory. Note that the SBBN predicted abundances are still subject to other uncertainties, such as the mean neutron lifetime $\tau_{\rm n}$ and nuclear reaction rates, but these values are measured in laboratories or inferred using \emph{ab initio} calculations. Primordial abundances deduced from observations of astrophysical regions thus provide a valuable test of the Standard Model of particle physics and cosmology and its assumptions.

Constraining the values of $\Omega_{\rm b} h^{2}$ and $N_{\rm eff}$ using observations requires using two or more measurements of the primordial abundances. For this exercise, we take our measurement of the primordial helium abundance in conjunction with the latest primordial deuterium abundance reported by \cite{2018ApJ...855..102C}:
\begin{gather*} 
Y_{\rm P}\,=\,0.2436^{+0.0039}_{-0.0040} \\
\rm (D/H)_{\rm P}\times10^{5}\,=\,2.527\pm0.030
\end{gather*}
and use calculations of BBN to infer the values of $\Omega_{\rm b} h^{2}$ and $N_{\rm eff}$ that best fit these abundances.

In what follows, we use the detailed primordial abundance calculations reported by \cite{2018PhR...754....1P}. 
These authors provide formulae for calculating the primordial abundances, given a value of $\Omega_{\rm b} h^{2}$, $N_{\nu}$, and $\tau_{\rm n}$. We restate the formula for predicting $Y_{\rm P}$ here as an example (see their Equation 145, the surrounding text, and Table VI of their paper for the values of the $C_{pqr}$ coefficients that are referenced here):
\begin{equation*}
\ \frac{\Delta Y_{\rm P}}{Y_{\rm P}} = \sum\limits_{pqr} C_{pqr}\Big(\frac{\Delta \Omega_{\rm b} h^{2}}{\Omega_{\rm b} h^{2}}\Big)^{p} \Big(\frac{\Delta N_{\nu}}{N_{\nu}}\Big)^{q} \Big(\frac{\Delta \tau_{\rm n}}{\tau_{\rm n}}\Big)^{r}
\end{equation*}

We use the latest measurement of the mean neutron lifetime $\tau_{\rm n}\,=\,877.7\pm0.7$ \citep{2018Sci...360..627P} to solve for $\Omega_{\rm b} h^{2}$ and $N_{\nu}$. Furthermore, we use the scaling $N_{\rm eff}\,=\,N_{\nu}\,\times\,3.046/3$ (see \citealt{2018PhR...754....1P}). This choice of scale is commonly used, and allows us to fairly compare the BBN results to the CMB. We use \textsc{emcee} with 100 walkers taking 1500 steps each and sample the parameter space:
\begin{gather*} 
0.0185 \leq \Omega_{\rm b} h^{2} \leq 0.0267 \\
1.5 \leq N_{\rm eff} \leq 4.5
\end{gather*}

\begin{figure}[!h]
\includegraphics[width=0.5\textwidth]{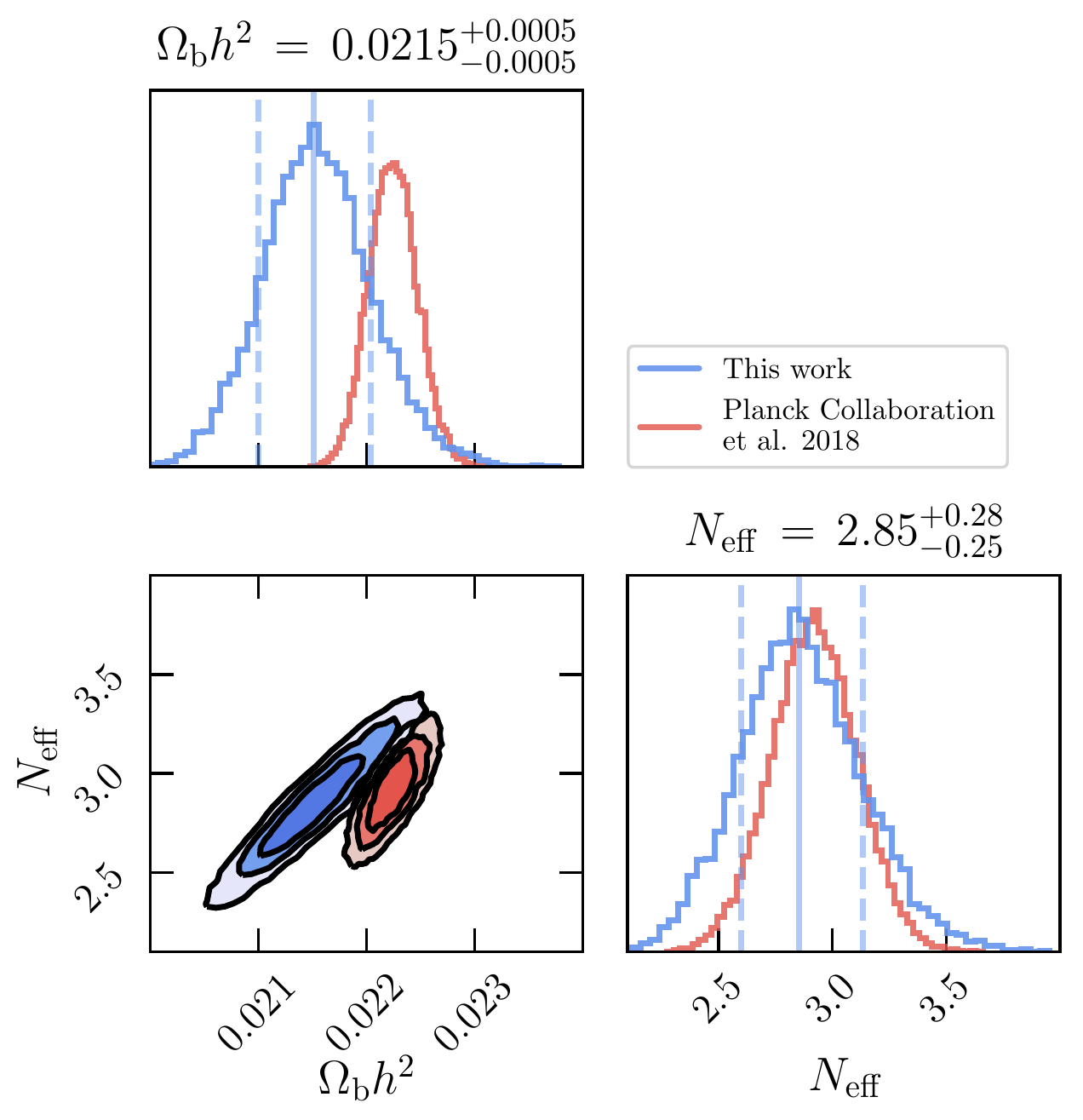}
\caption{The results of the MCMC analysis performed to recover the most likely values of $\Omega_{\rm b} h^{2}$ and $N_{\rm eff}$, given our latest primordial helium abundance measurement and the \cite{2018ApJ...855..102C} primordial deuterium abundance (blue contours and histograms). The quoted values above each histogram are as recovered via our analysis. The blue solid line in the histogram indicates the best recovered value, and the blue dashed lines show the 1$\sigma$ bounds. The red contours and histograms show the constraints on $\Omega_{\rm b} h^{2}$ and $N_{\rm eff}$ as measured by the \textit{Planck} satellite \citep{2018arXiv180706209P}. The contours show the 1$\sigma$, 2$\sigma$, and 3$\sigma$ levels.}
\label{fig:BBN_param_estimation}
\end{figure}

With each step, a model set of primordial $\rm D/\rm H$ and $Y_{\rm P}$ abundances are predicted. The optimal parameters are solved for assuming a Gaussian likelihood function. We take the burn in to be at 0.8$\times\,n_{\rm steps}\,=\,$ 1200 steps. Given the observed primordial element abundances, we report the following bounds on the effective number of neutrino species and the baryon density:
\begin{gather*} 
N_{\rm eff}\,=\,2.85^{+0.28}_{-0.25} \\
\Omega_{\rm b} h^{2}\,=\,0.0215^{+0.0005}_{-0.0005}
\end{gather*}
\noindent The result of our MCMC calculation is shown in Figure \ref{fig:BBN_param_estimation}, together with the \textit{Planck} bounds on these parameters.\footnote{We use the \textit{Planck} Release 3 data with the prefix $`$plikHM$\_$TTTEEE$\_$lowl$\_$lowE'.} The best fit value of $N_{\rm eff}$ is consistent with the value inferred by \textit{Planck} ($N_{\rm eff}\,=\,2.92^{+0.19}_{-0.18}$; shown by the red contours in Figure \ref{fig:BBN_param_estimation}) and the Standard Model value of $N_{\rm eff}\,=\,3.046$.

\subsection{Future Improvements}
\label{future_improvements}
The goal of the spectroscopic survey reported by \cite{2018ApJ...863..134H} was twofold: (1) to increase the number of known systems in the low-metallicity regime and in particular, push on the lowest-metallicity regime, and (2) to obtain high-quality optical and NIR spectroscopy of a subset of the new, metal-poor galaxies, with priority on systems with metallicities determined to be $12+\textnormal{log}_{10}(\rm O/\rm H)\,\leq\,7.65$ based on strong-line calibration methods. The purpose of the specific goals was to better populate and constrain the metal-poor end in the extrapolation to a primordial helium abundance. In the following text, we discuss the current limitations of the PHLEK survey and future improvements that need to be explored to push $y_{\rm P}$ to sub-percent level accuracy.
\subsubsection{Qualification Rates}
\label{future_observing}
One of the main obstacles of measuring the primordial helium abundance is the difficulty of accurately modelling a large fraction of emission line observations. To give an overview of the current status of modelling the \HII\ region emission lines, in Table \ref{table:qualify_vs_nonqualify} we have compiled the qualification rates of the three survey samples considered in this paper. Our results show that the qualification rates of \HII\ regions that only have optical data are consistent among the PHLEK, SDSS, and HeBCD data sets -- about 10$\%$ of systems make it into Sample 1. The meager number of currently known, near-pristine systems that push on the lowest-metallicity regime hinders our ability to constrain the slope (and thus intercept) of the linear extrapolation to the primordial value. The effect of a dearth of the most metal-poor systems is multiplied when these systems are unsuccessfully modelled and consequently excluded from primordial helium analyses after quality screening. 

However, Table \ref{table:qualify_vs_nonqualify} shows that systems with complementary optical and NIR data are more successfully modelled. Our PHLEK sample sees an increase from 10$\%$ to 13$\%$ of systems qualifying for Sample 1 and more noticeably, the HeBCD sample success rate increases to 29$\%$ when NIR data are included. An aspect that contributes to the small fraction of systems that can be well-modelled lies in the difficulty of confidently detecting the weak optical \HeI\ lines necessary for accurately determining the physical conditions of the \HII\ region, including the helium abundance. The addition of the NIR \HeI~$\lambda$10830 line to primordial helium work saw an appreciable reduction in the errors on the recovered helium abundances and electron densities \citep{2014MNRAS.445..778I, 2015JCAP...07..011A}. The value of the \HeI~$\lambda$10830 line is likely the reason behind the higher success rates we see in Table \ref{table:qualify_vs_nonqualify} for systems with optical and NIR spectroscopy. The sensitivity of the \HeI~$\lambda$10830 line emissivity to the electron density eliminates degeneracies between the electron temperature and electron density when modelling systems. Although the \HeI~$\lambda$10830 line is the brightest emission line detected in our NIR observations, it still sometimes eluded detection completely. From our experience, it is useful to target systems with $F$(H$\beta)\,\gtrsim~10^{-15}$\,erg\,s$^{-1}$\,cm$^{-2}$ for complementary NIR spectroscopy. This assumes existing optical data, but the criteria increases the chance of acquiring high $S/N$ NIR data. 

Even with our systems that satisfy this criteria, however, we recover a lower success rate in modelling the PHLEK sample compared to the HeBCD sample. Table \ref{table:qualify_vs_nonqualify} shows that a total of 27$\%$ of our systems with optical plus NIR spectroscopy qualify in Sample 2 (which includes the systems that qualify in Sample 1), compared to 67$\%$ for the HeBCD sample. We presume the difference comes from these two data sets consisting of different types of galaxies. The correlation between $F$(H$\beta$) and a NIR detection mentioned above is not unlike the criteria imposed by \citeauthor{2004ApJ...602..200I} as part of their HeBCD sample selection. Specifically, the construction of the HeBCD data set was based on existing observations and used a selection criteria of high $EW$(H$\beta$), quoted to be generally $EW$(H$\beta)\,\geq\,200\,$\AA, and have metallicities ranging from $12+\textnormal{log}_{10}(\rm O/\rm H)\,=\,7.00-8.21$ (\cite{2004ApJ...602..200I}; although we note that only 35 of the 93 HeBCD sample satisfies the $EW(\rm H\beta$) condition, and 22 of the 93 fall in the low-metallicity regime). Subsequent analysis of the HeBCD data set by \cite{2007ApJ...662...15I} was combined with SDSS DR5 spectroscopy, with the requirement that only SDSS galaxies with $EW$(H$\beta)\,\geq\,50\,$\AA, and $F$(H$\beta)\,\gtrsim~10^{-14}$\,erg\,s$^{-1}$\,cm$^{-2}$ are included in the analysis. While we also impose a $EW$(H$\beta)\,\geq\,50\,$\AA\, criteria (see Section \ref{qualification}), a comparison of the $F$(\HeI~$\lambda$10830)\,/\,$F$(P$\gamma$) detection levels shows that all HeBCD systems have \HeI~$\lambda$10830\AA\, to P$\gamma$ ratios detected with $S/N\,\geq\,50$, whereas the PHLEK sample have $S/N\,\leq\,50$, regardless of the measured $EW$(H$\beta$) and the status of the modelling success rate.

Thus, it is evident that a sample selection based on measured high $EW$(H$\beta$), as with the HeBCD data set, versus one based on low estimated metallicities via strong-line calibrations, as with the PHLEK sample, yield different data sets. The former yielded systems with higher significance detections of the necessary NIR \HeI~$\lambda$10830 emission line and a higher modelling success rate. Meanwhile, the latter successfully populates the lowest-metallicity end of the galaxy sample, but currently face more significant limitations in accurately modelling their physical conditions and characteristics, even with \HeI~$\lambda$10830. Until we identify and improve current shortcomings, which may lie in data processing or in model simplicities and deficiencies, we are not equipped to equally include all types of metal-poor systems to deduce the primordial helium abundance. Now that a substantial sample of metal poor star-forming galaxies are known, we suggest that a more detailed analysis of individual systems may allow us to better model the complicated physics of \HII\ regions. This in turn may allow us to construct a model with improved capabilities of recovering the helium abundance in a wider variety of star-forming galaxies.

\begin{deluxetable*}{lccc c ccc}[ht!]
\tablecaption{Success Rates for Modelling our Dataset} 
\tablehead{ 
 & \multicolumn{3}{c}{Optical+NIR} & & \multicolumn{3}{c}{Optical Only} \\
\cline{2-4} \cline{6-8}
\colhead{Data Set} & \colhead{Sample 1} & \colhead{Sample 2} & \colhead{Total Systems} & & \colhead{Sample 1} & \colhead{Sample 2} & \colhead{Total Systems} \\
\colhead{} & \colhead{Number ($\%$)} & \colhead{Number ($\%$)} & \colhead{} & & \colhead{Number ($\%$)} & \colhead{Number ($\%$)} & \colhead{}
}
\startdata 
PHLEK & 2 (13\%) & 4 (27\%) & 15 & & 1 (10\%) & 3 (30\%) & 10 \\
SDSS & \nodata & \nodata & \nodata & & 38 (7\%) & 85 (15\%) & 578 \\
HeBCD & 6 (29\%) & 14 (67\%) & 21 & & 7 (10\%) & 12 (17\%) & 69 \\
\enddata
\tablecomments{The number (and percentage) of systems from our PHLEK sample, the SDSS sample, and the HeBCD sample that qualify for Sample 1 and Sample 2, out of the total number of systems available in each sample. The statistics are separated by systems for which optical and NIR spectroscopy are available and systems for which only optical spectroscopy exists. We remind readers that Sample 1 is included in Sample 2, and the two are described in Sections \ref{sample1} and \ref{sample2} respectively. The HeBCD optical+NIR sample has the highest rate of satisfying our criteria for Sample 1 and Sample 2, likely due to the higher $S/N$ NIR data that exist for the HeBCD sample. The qualification rate for systems with only optical data are comparable across all data sets and lower than the rate for systems with complementary NIR data in the same data set when the comparison is available. These rates illustrate the difficulty of modelling systems well without complementary NIR spectroscopy and more importantly, the need for high-quality NIR data.}
\label{table:qualify_vs_nonqualify}
\end{deluxetable*}

\subsubsection{Towards a Sub-percent Measurement of $y_{\rm P}$}
\label{future_model}
Other potential obstacles faced by primordial helium analyses come post-data collection, a facet of which is in data processing. Currently, fluxing our emission line spectra using observations of spectrophotometric standards introduces uncertainties in \emph{relative} flux measurements between 1--2$\%$, shown by \cite{1990AJ.....99.1621O}. The weakest \HeI\ lines and their measured flux ratios are therefore easily susceptible to errors introduced during flux calibration. To push observational primordial helium measurements to the sub-percent level will require flux calibrations to mirror this precision. \cite{2004ApJ...602..200I, 2007ApJ...662...15I} take precaution to derive sensitivity curves using only hot white dwarf standard stars that show relatively weak absorption features, such as Feige 34, Feige 110, and HZ 44. Such stars allow for sensitivity curves that are accurate to $\lesssim1\%$ over the optical wavelength range \citep{1990AJ.....99.1621O}. Following this necessity of sub-percent flux calibrations, we propose the use of the near-perfect blackbody stars from \cite{2018AJ....156..219S} to flux calibrate observations of metal-poor \HII\ regions. The spectra of these stars, thought to be white dwarfs, are nearly featureless, and their blackbody nature ranges from the ultraviolet to infrared. These stars offer the potential to improve the precision of flux calibrations even further and can bring primordial helium abundances closer to the sub-percent level.

Additionally, the model we assume in this work to describe our emission line observations is subject to deficiencies. As part of our analysis, we investigated obvious shortcomings in our model, such as the inability to model a specific emission line or the inability to model systems when their parameters fall in a particular regime. Reassuringly, we found no obvious parameters or combination of parameters that perform poorly for non-qualifying systems. However, since the model is unable to reproduce a high fraction of the initial galaxy sample, we conclude that some aspects of \HII\ region modelling are currently unaccounted for.

Helium abundance measurements have historically been derived from longslit observations. These observations are assumed to be representative of the entire \HII\ region. In reality, the integrated light that enters the slit likely samples multiple radii of an \HII\ region and can also be the result of multiple, overlapping \HII\ regions. Such simplistic assumptions likely affect our ability to fit the observed data with a single set of parameters. Possible model enhancements include dropping the simple two-zone photoionization model characterized by two temperatures, and introducing a temperature structure to our model. However, we note that we do not anticipate the temperature structure within an \HII\ region to vary much beyond the limits we place on our temperature prior (selected to be $\sigma\,=\,0.2T_{\rm m}$, where $T_{\rm m}$ is the direct measurement of the electron temperature from the [\OIII] lines). Similarly, the density of the \HII\ region is likely a function of distance from the central star, and introducing a density structure may improve our model as well.




\section{Summary and Conclusion}
\label{conclusion}
We present a sample of NIR observations of several metal-poor galaxies reported by \cite{2018ApJ...863..134H}. Using this sample along with galaxies from the SDSS spectroscopic database and existing metal-poor galaxies in the literature, we report a new determination of the primordial helium abundance. We summarize the main results of our analysis as follows: 

\begin{enumerate}
\item We obtain near-infrared (NIR) spectra of a sample of sixteen galaxies to complement optical spectroscopy presented by \cite{2018ApJ...863..134H}. The NIR observations are taken using NIRSPEC or NIRES at Keck Observatory and are designed to obtain a measurement of the \HeI~$\lambda$10830\AA\, to P$\gamma$ flux ratio. We supplement this sample with 1053 starburst galaxies in the SDSS spectroscopic database, selected based on their star-forming nature and sufficiently high $S/N$ data on a suite of optical \HeI\ lines, and 93 systems from the \cite{2007ApJ...662...15I} HeBCD sample, a subset of which include follow-up NIR observations reported by \cite{2014MNRAS.445..778I}. 

\item We outline our Python-based code \textsc{yMCMC}, which uses a Markov Chain Monte Carlo (MCMC) approach to find parameters that best describe the observed emission line flux ratios of each galaxy. The parameters we solve for in the analysis are: the singly ionized helium abundance, the electron temperature and density, the reddening parameter, the underlying \HI\ and \HeI\ stellar absorption, the helium optical depth parameter, and the ratio of neutral to ionized hydrogen densities. Our method is largely based on the approach developed by \cite{2011JCAP...03..043A, 2012JCAP...04..004A, 2013JCAP...11..017A, 2015JCAP...07..011A}. Our implementation of the techniques include new \HI\ emissivities that extend to the lowest density regime, $n_{\rm e}\,=\,1~\rm cm^{-3}$, a different treatment of the blended H8+\HeI~$\lambda$3889 lines, and a different method of correcting for \HI\ emission stemming from collisional excitation. We also employ an alternative qualification approach, which requires that we statistically reproduce \emph{all} emission lines used in the analysis.

\item Using \textsc{yMCMC}, we solve for the best fit parameters that describe our sample of galaxies. 
We construct two qualifying samples, named ``Sample 1'' and ``Sample 2''. Sample 1 contains all galaxies whose \HI\ and \HeI\ emission lines are reproduced by our model to within 2$\sigma$. Sample 2 is defined such that all except one observed emission line ratios are reproduced to within 2$\sigma$ (however, the one emission line that fails this 2$\sigma$ limit must be reproduced to within 3$\sigma$).

\item We calculate ionic abundances of O$^{+}$, O$^{++}$, and $y^{++}$, and combine these with the $y^{+}$ values recovered from the MCMC analysis to calculate total number abundance ratios, $\rm O/\rm H$ and $y$. We fit a linear model to the $\rm O/\rm H$ versus $y$ abundances of Sample 1 and Sample 2, and extrapolate to zero metallicity to infer the primordial helium abundance, $y_{\rm P}$. Our linear model allows for the presence of an intrinsic scatter of the measurements due to systematic uncertainties that may currently be unaccounted for. We find that Sample 1 contains no evidence of intrinsic scatter, while Sample 2 contains some intrinsic scatter. Both samples yield primordial helium abundances that are in mutual agreement with one another. However, we adopt the $y_{\rm P}$ determination based on Sample 1 due to our increased confidence in the model. We report a primordial helium number abundance ratio $y_{\rm P}\,=\,0.0805^{+0.0017}_{-0.0017}$, which corresponds to a primordial helium mass fraction $Y_{\rm P}\,=\,0.2436^{+0.0039}_{-0.0040}$.

\item Combining our determination of $y_{\rm P}$ with $(\rm D/\rm H)_{\rm P}$ from \cite{2018ApJ...855..102C}, we find $\Omega_{\rm b} h^{2}\,=\,0.0215^{+0.0005}_{-0.0005}$ and $N_{\rm eff}\,=\,2.85^{+0.28}_{-0.25}$. This value of $N_{\rm eff}$ is within 1$\sigma$ agreement with the Standard Model value of $N_{\rm eff}\,=\,3.046$. Our value of $\Omega_{\rm b} h^{2}$ is in 1.3$\sigma$ agreement with the value measured by \textit{Planck}.
\end{enumerate}

Observational measurements of the primordial light element abundances provide a unique window to study the conditions of the early Universe and offer the potential to identify non-standard physics at the time of BBN. The latest $(\rm D/\rm H)_{\rm P}$ determination has reached the percent level, comparable to the precision achieved by the latest CMB constraints. $y_{\rm P}$ determinations are reaching similar precision; in particular, the recent addition of the NIR \HeI~$\lambda$10830 line to helium abundance analyses has led to an improvement in the helium abundance measurements of individual galaxies. 

From a theoretical perspective, the primordial helium abundance can be reliably calculated; a high precision observational determination of the primordial helium abundance will therefore provide the most sensitive test of the Standard Model. As we move towards this era of high precision cosmology, we advocate that it will become necessary to understand the nuances of our current limitations of \HII\ region modelling and push data processing techniques to higher accuracy. These can potentially be done by conducting detailed observations of individual systems and improving flux calibration using near-featureless blackbody stars as standards. Finally, we strongly suggest that observational primordial helium works, and also BBN calculations, shift towards reporting helium \emph{number} abundances, as opposed to a helium mass fraction commonly adopted in literature, for the most direct comparison to theoretical works.

\acknowledgments
We are grateful to the anonymous referee for their thorough review and helpful comments, which have resulted in an improved manuscript. The authors are indebted to Erik Aver for providing fluxes and equivalent widths from their 2015 work and generating synthetic fluxes, both of which were extremely valuable to the development and testing of our code. We also thank Evan Skillman for insightful discussions on the current limitations of primordial helium analyses and on areas for future improvement. We are grateful to Peter Storey for providing calculations of the latest hydrogen emissivities down to the low densities that characterize \HII\ regions.

The data presented herein were obtained at the W.M. Keck Observatory, which is operated as a scientific partnership among the California Institute of Technology, the University of California and the National Aeronautics and Space Administration. The Observatory was made possible by the generous financial support of the W.M. Keck Foundation. The authors wish to recognize and acknowledge the very significant cultural role and reverence that the summit of Mauna Kea has always had within the indigenous Hawaiian community.  We are most fortunate to have the opportunity to conduct observations from this mountain. We gratefully acknowledge the support of the staff at Keck Observatory for their assistance during our observing runs. During this work, R.~J.~C. was supported by the Royal Society University Research Fellowship, UF150281. R.~J.~C. acknowledges support from STFC (ST/L00075X/1, ST/P000541/1). J.~X.~P. acknowledges support from the National Science Foundation grant AST-1412981.

\facilities{Keck:I (LRIS), Keck:II (NIRSPEC, NIRES)}
\software{\texttt{astropy} \citep{2013A&A...558A..33A},\, \texttt{matplotlib} \citep{Hunter:2007},\, \texttt{NumPy}\,\citep{NumPy},\,\texttt{SciPy}\, \citep{SciPy}, \texttt{PypeIt}, \citep{j_xavier_prochaska_2019_3506873}}

\appendix
\section{SDSS CasJobs Query}
\label{CasJobsquery}
The following text shows the CasJobs query submitted to retrieve the full sample of galaxies in the SDSS spectroscopic database potentially suited to be included in our primordial helium abundance determination. The query requires that these systems be star-forming galaxies within a redshift range of $0.02\,<\,z\,<\,0.15$. This guarantees that the [\OII] doublet and \HeI~$\lambda$7065 lines, necessary for oxygen and helium abundance measurements, are detected.
\begin{center}
\texttt{SELECT} e.specObjID, e.ra, e.dec, e.z, e.zErr, e.subClass, \\ dbo.fGetUrlFitsSpectrum(e.specObjID) as urlfits \texttt{into} mydb.specobj\_starburst \\
\texttt{FROM} specObj AS e \\
  \texttt{WHERE} e.z < 0.15 \\
  \texttt{AND} e.z > 0.02 \\
  \texttt{AND} e.subClass = 'STARBURST' \\
\end{center}

\vspace{1cm}

\section{Mock Data and MCMC Recovery}
\label{mock_mcmc}
Given a set of input parameter values and $EW$s, we generate mock data (i.e., the would-be observed flux ratios) to test how well \textsc{yMCMC} can recover the input parameters. In all these test runs, we adopt a weak temperature prior equal to the input temperature of $T_{\rm e}$\,=\,18,000~K. The results are shown in the following table.

\begin{deluxetable}{cc cc cc}[h!]
\tablecaption{MCMC recovery on mock data} 
\tablehead{ \colhead{} & \colhead{} & \multicolumn{4}{c}{Recovered Values} \\ 
\cline{3-4} \cline{5-6} 
\colhead{} & \colhead{} & \multicolumn{2}{c}{Optical+NIR} & \multicolumn{2}{c}{Optical Only} \\
\colhead{Parameter} & \colhead{Input Value} & \colhead{All} & \colhead{no $\frac{F(\rm H\alpha)}{F(\rm H\beta)}$} & \colhead{All} & \colhead{no $\frac{F(\rm H\alpha)}{F(\rm H\beta)}$}
}
\startdata 
$y$+ & 0.0800 & 0.0801$^{+0.0047}_{-0.0043}$ & 0.0809$^{+0.0051}_{-0.0049}$ & 0.0800$^{+0.0047}_{-0.0051}$ & 0.0816$^{+0.0050}_{-0.0054}$ \\
$T_{\rm e}$ & 18000 & 18000$^{+2100}_{-2000}$ & 17700$^{+2000}_{-2000}$ & 18000$^{+2100}_{-2200}$ & 17800$^{+2000}_{-2200}$ \\
log$_{10}(n_{\rm e}/\rm cm^{-3}$) & 2.0 & 2.00$^{+0.15}_{-0.16}$ & 1.36$^{+0.78}_{-0.92}$ & 2.00$^{+0.16}_{-0.16}$ & 1.56$^{+0.65}_{-1.03}$ \\
$c$(H$\beta$) & 0.1 & 0.010$^{+0.01}_{-0.01}$ & 0.111$^{+0.02}_{-0.02}$ & 0.101$^{+0.02}_{-0.02}$ & 0.11$^{+0.02}_{-0.03}$ \\
$a_{\rm H}$ & 1.0 & 0.93$^{+0.31}_{-0.33}$ & 0.68$^{+0.49}_{-0.37}$ & 0.89$^{+0.42}_{-0.40}$ & 0.66$^{+0.58}_{-0.42}$ \\
$a_{\rm He}$ & 1.0 & 1.01$^{+0.37}_{-0.38}$ & 1.01$^{+0.40}_{-0.39}$ & 1.0$^{+0.38}_{-0.36}$ & 1.06$^{+0.39}_{-0.41}$ \\
$\tau_{\rm He}$ & 1.0 & 1.01$^{+0.46}_{-0.41}$ & 1.24$^{+0.61}_{-0.56}$ & 1.011$^{+0.50}_{-0.41}$ & 1.19$^{+0.62}_{-0.55}$ \\
log$_{10}(\xi$) & -4.0 & $\leq$ -2.73 & $\leq$ -2.74 & $\leq$ -2.68 & $\leq$ -2.66 \\
\enddata
\tablecomments{MCMC results on the recovery of parameters, given mock data. The corresponding contours and histograms showing the best recovered model parameters on these trial runs are shown in Figures \ref{fig:mock_nir}, \ref{fig:mock_opt}, \ref{fig:mock_nir_noHaHb}, and \ref{fig:mock_opt_noHaHb}.}
\label{table:mcmc_recovery_tests}
\end{deluxetable}

\newpage

\begin{figure}[h!]
\includegraphics[width=1.0\textwidth]{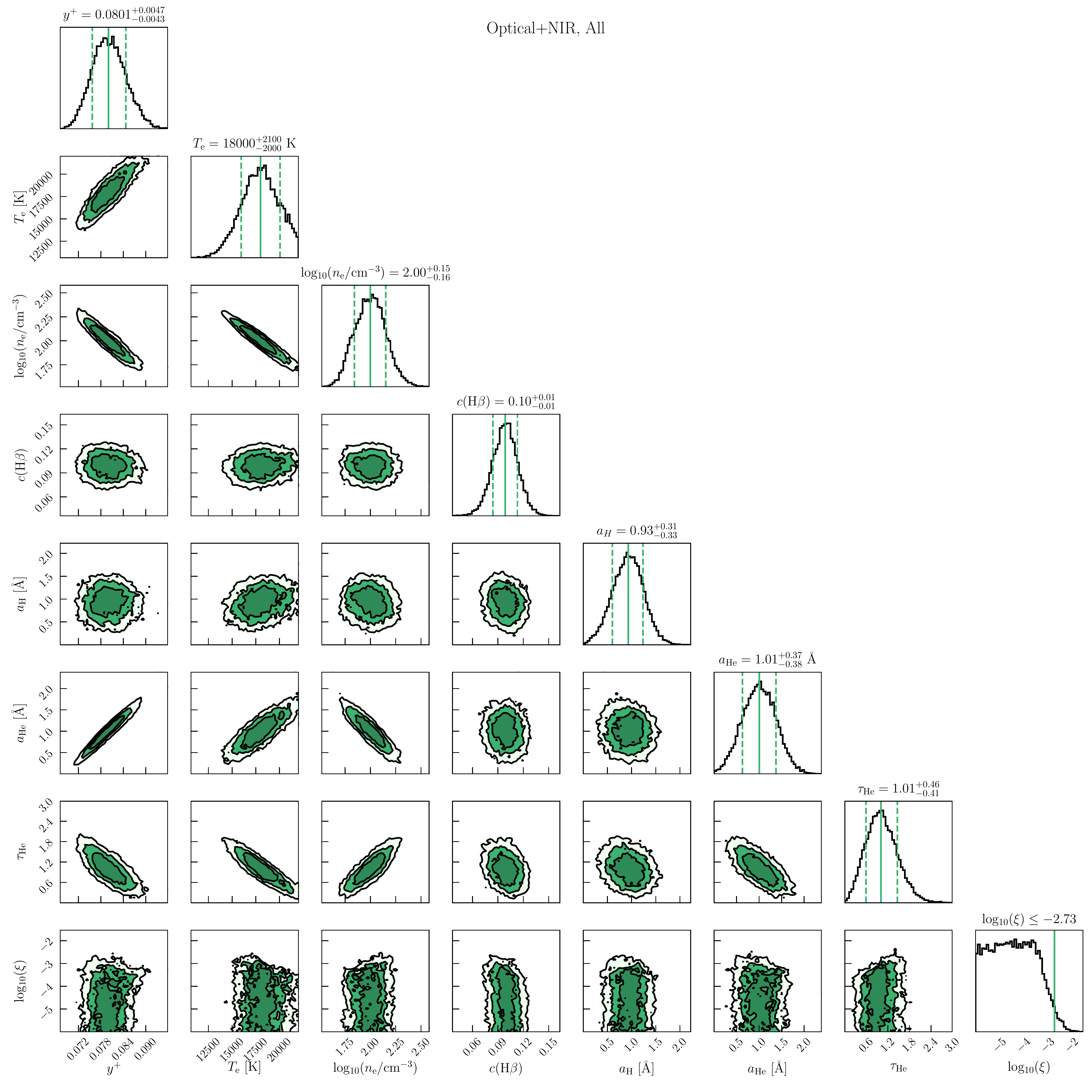}
\caption{Contours (off-diagonal panels) and histograms (diagonal panels) showing the best recovered model parameters on mock data including optical and near-infrared data, with the $F(\rm H\alpha)/F(\rm H\beta)$ ratio, i.e., the first column of recovered values in Table \ref{table:mcmc_recovery_tests}. The contours show the 1$\sigma$, 2$\sigma$, and 3$\sigma$ levels. The solid green line in the histograms show the best recovered parameter value, and the dotted green lines show the $\pm1\sigma$ values. In the panel showing the results for the log$_{10}(\xi)$ parameter, the solid vertical line represents a $2\sigma$ upper limit.}
\label{fig:mock_nir}
\end{figure}

\begin{figure}[h!]
\includegraphics[width=1.0\textwidth]{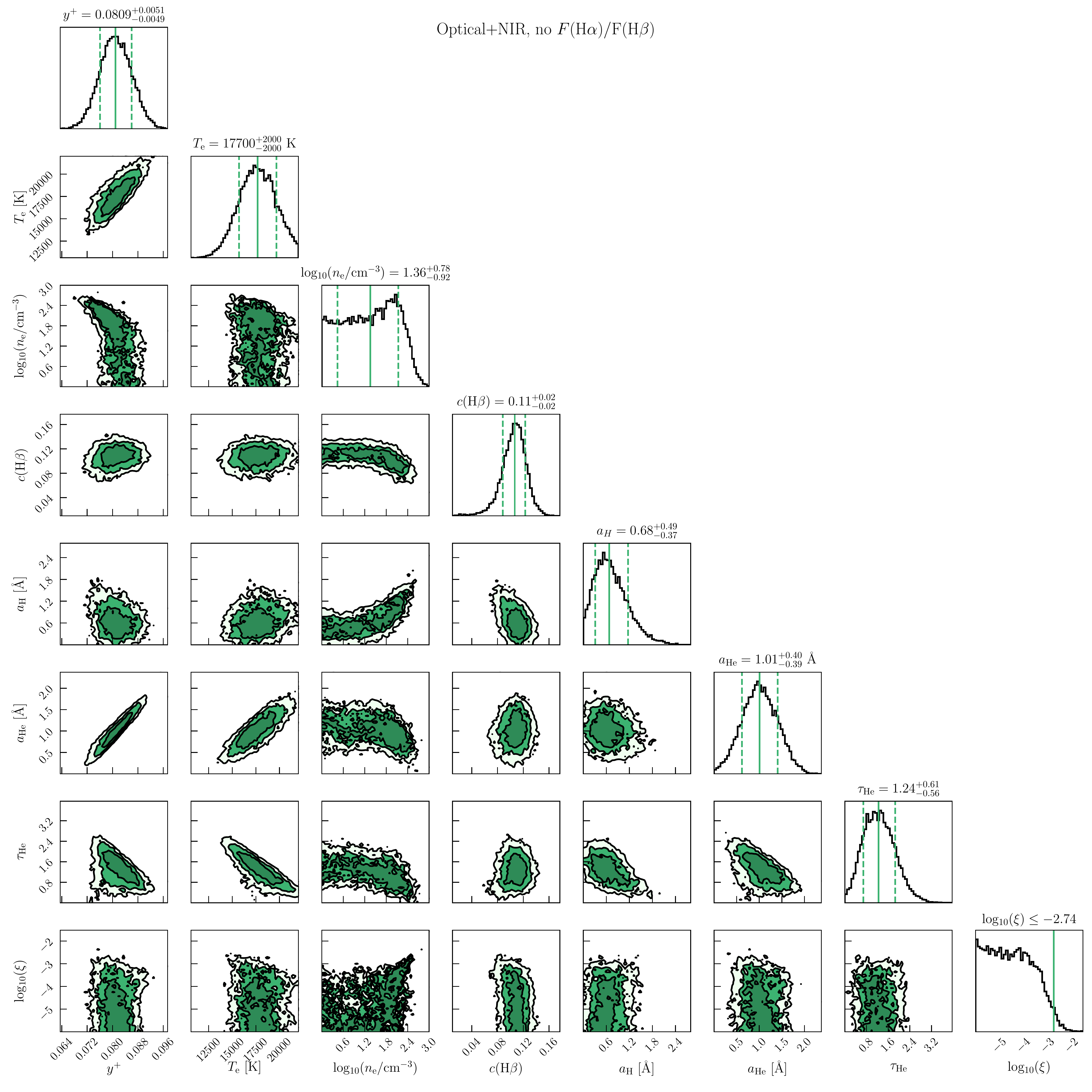} 
\caption{Same as Figure \ref{fig:mock_nir}, but on mock data including optical and near-infrared data, without the $F(\rm H\alpha)/F(\rm H\beta)$ ratio, i.e., the second column of recovered values in Table \ref{table:mcmc_recovery_tests}.}
\label{fig:mock_opt}
\end{figure}

\begin{figure}[h!]
\includegraphics[width=1.0\textwidth]{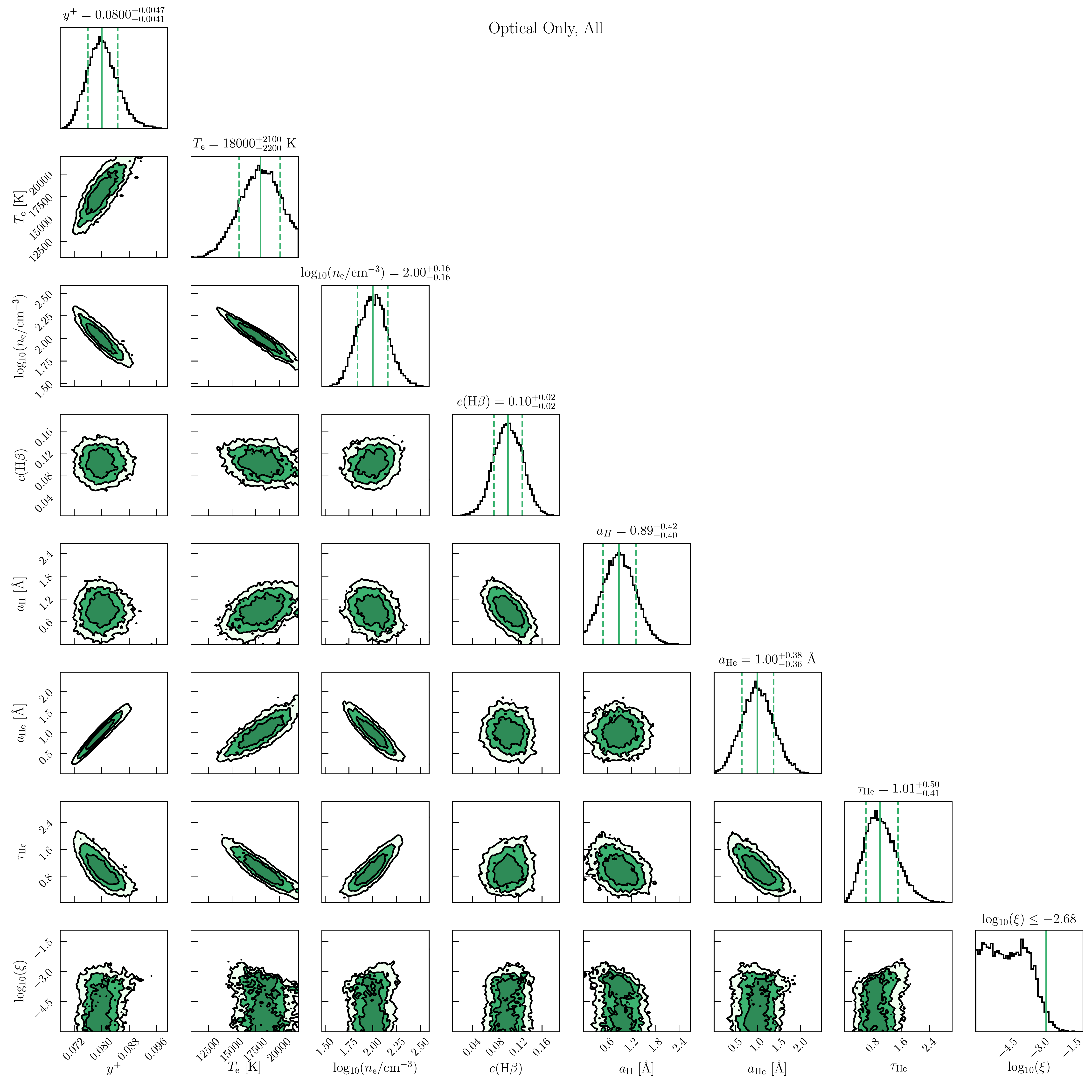}
\caption{Same as Figure \ref{fig:mock_nir}, but on mock data including only optical data, with the $F(\rm H\alpha)/F(\rm H\beta)$ ratio, i.e., the third column of recovered values in Table \ref{table:mcmc_recovery_tests}.}
\label{fig:mock_nir_noHaHb}
\end{figure}

\begin{figure}[h!]
\includegraphics[width=1.0\textwidth]{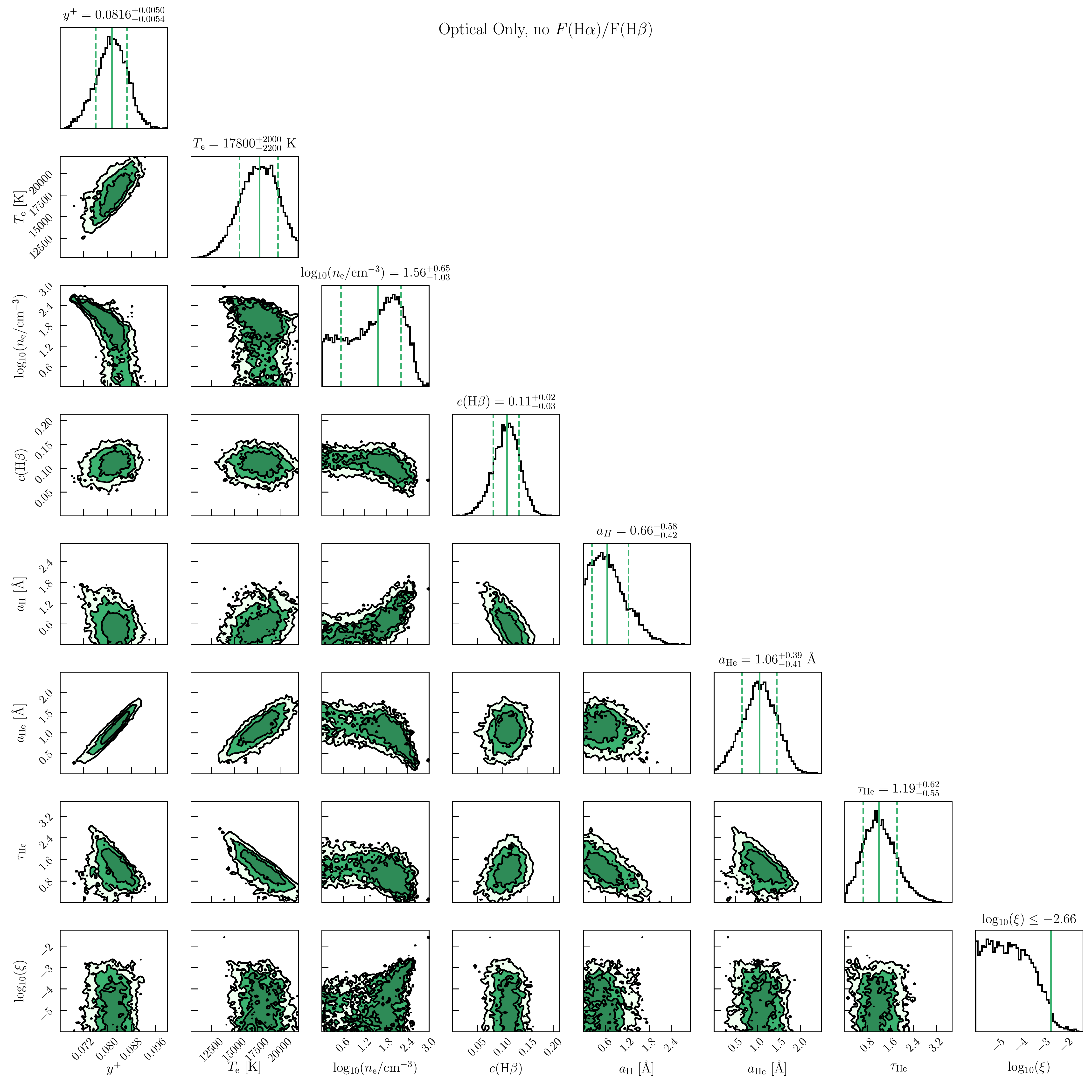}
\caption{Same as Figure \ref{fig:mock_nir}, but on mock data including only optical data, without the $F(\rm H\alpha)/F(\rm H\beta)$ ratio, i.e., the fourth column of recovered values in Table \ref{table:mcmc_recovery_tests}.}
\label{fig:mock_opt_noHaHb}
\end{figure}

\end{document}